\documentclass[tighten, times, preprint2]{aastex6} 

\providecommand{\adsurl}[1]{\href{#1}{ADS}}

\newcommand\tnm[1]{\tablenotemark{#1}}
\usepackage{amsmath}
\usepackage{hyperref}
\usepackage{threeparttable}

\shorttitle{Quantitative Classification of Type I Supernovae}
\shortauthors{Sun \& Gal-Yam }

\begin{document}

\title{Quantitative Classification of Type I Supernovae Using Spectroscopic Features at Maximum Brightness}

\author{Fengwu Sun\altaffilmark{1},
\and
Avishay Gal-Yam\altaffilmark{2}
}

\affiliation{\altaffilmark{1} Department of Astronomy, Peking University, Beijing 100871, China \\
       \altaffilmark{2} Department of Particle Physics and Astrophysics, Weizmann Institute of Science, Rehovot, Israel; 
       \href{mailto:avishay.gal-yam@weizmann.ac.il}{avishay.gal-yam@weizmann.ac.il}}

\begin{abstract}
We present a set of new quantitative classification criteria for major subclasses of Type I Supernovae (SNe). We analyze peak spectra of 146 SNe Ia from the Berkeley Supernova Ia Program (BSNIP), 12 SNe Ib, 19 SNe Ic (including 5 SNe Ic-BL) and 4 SNe Ib/c. All the spectra are within 5 days of maximum light. We study their absorption line depths relative to the pseudo-continuum at around observed wavelength $\lambda$6150\AA, attributed to Si II $\lambda$6355\AA \ for SNe Ia and Ic and to hydrogen for SNe Ib, and O I $\lambda 7774$\AA\ observed around $\lambda 7500$\AA. We found that Type I SNe could be quantitatively classified using the line depths at these two absorption regions. Type Ia SNe, including normal SNe Ia, Ia-1991bg and Ia-1999aa, show strong signatures of Si II near peak with relative line depth values of $a{\rm (\lambda6150\AA)>0.35}$. Only SNe Ia-2002cx (Iax) do not satisfy this criterion, while the $a{\rm (O\ I\ \lambda 7774\AA)}$ index can separate SNe Ia-1991bg and Ia-1999aa cleanly.
 Type Ib SNe satisfy that $a{\rm (\lambda6150\AA)<0.35}$ and $a{\rm (\lambda6150\AA)} / a{\rm (O\ I\ \lambda 7774\AA)>1}$, while regular Type Ic SNe show $a{\rm (\lambda6150\AA)<0.35}$ but $a{\rm (\lambda6150\AA)} / a{\rm (O\ I\ \lambda 7774\AA)<1}$. These quantitive classification criteria work well for the majority of normal Type I SN subclasses. Peculiar SNe Ia-2002cx and SNe Ic-BL are the exception. We apply these criteria to provide clear identifications to intermediate or unclear SNe Ib/c. The results show good consistency with classification using specific spectral line features. Though there are still some puzzles about the physical mechanisms driving the diverse line ratios, our method is an accurate empirical method to identify the majority of Type I SNe. 
\end{abstract}


\section{INTRODUCTION}
\label{introduction}
The classification of SNe has been a systematic and long-lasting problem for astronomers since the middle of the 20th century. The first classification of SNe, based on spectroscopic data, was introduced by \citet{1941PASP...53..224M}. Minkowski's well-known classification criterion is that those SNe whose spectra show prominent signatures of Hydrogen were classified as Type II SNe, while Type I SNe did not show evidence for Hydrogen in their spectra.

After this first influential step, the following decades witnessed the introduction of several new SN subclasses. 
Among the Hydrogen-deficient Type I SNe, a new subclass of SNe Ib was noticed and differentiated from the dominant group (SNe Ia) by \citet{1985ApJ...294L..17W} and \citet{1985ApJ...296..379E}. 
Later \citet{1987ApJ...317..355H} identified signatures of He I in peak spectra of SNe Ib and \citet{1990RPPh...53.1467W} introduced the term SN Ic for a subclass of SNe Ib that did not show strong Helium absorption but were otherwise similar to SNe Ib (and different from SNe Ia). 

During the last 20 years, with the rapid development of observation and data reduction techniques as well as the launch of several ambitious SN survey projects, a number of Type I SNe with peculiar photometric properties or spectroscopic features have been discovered, leading to the establishment of several subclasses, 
for instance, SN Ia 1991bg-like events whose prototype was first described by \citet{1992AJ....104.1543F}, SN Ia 2002cx-like events, initially presented by \citet{2003PASP..115..453L} and discussed in detail in \citet{2013ApJ...767...57F} and \citet{White2015}, as well as broad-line Type Ic SN events (Ic-BL), showing extreme values of velocity dispersion at and before peak which cause individual spectral lines to blend together \citep[e.g.][]{Mazzali2000,2013ApJ...776...98X,2016ApJ...832..108M}. Some SNe Ic-BL events attract a lot of attention due to their association with high-energy GRB events \citep[for a review see][]{2006ARA&A..44..507W}. All of these discoveries enhance the diversity of Type I SNe, while a connection with the physical mechanism of SN explosion remains elusive. An updated review of SN classification is presented in \citet{2016arXiv161109353G}.

The studies of \citet{2011Natur.480..344N} and \citet{2012ApJ...744L..17B} of the SN Ia 2011fe constrain the size of its progenitor, making a compact white dwarf (WD) the most likely progenitor star. 
However, the detailed physical scenario of these explosive events still remains an open question. 
Meanwhile, we have strong evidence that normal Type Ib and Ic SNe originate from massive progenitor stars \citep[e.g.][]{van1992,Smartt2009}. 
The study of the SN Ib iPTF13bvn pre-explosion imaging, presented by \citet{2013ApJ...775L...7C} as well as \citet{Fremling2014}, shows an apparent massive progenitor and directly demonstrates the physical origin of SNe Ib.
Direct evidence of the progenitors of SNe Ic is still missing. From similar spectral features shared with SNe Ib as well as host galaxy population and locations, it is widely believed that SNe Ic are also the result of the core collapse of massive stars. 

Historically, SNe are classified by spectroscopic features \citep{Alex1997,2016arXiv161109353G}. For example, SN Ia spectra around maximum light are characterized by conspicuous absorption around $\lambda$6150\AA , which is produced by blueshifted Si II $\lambda$6355\AA, while this feature is weak in SN Ib and Ic spectra \citep{1995PhR...256..211W}.
SNe Ib are defined by displaying strong He I $\lambda\lambda\lambda$5876\AA\ 6678\AA\ 7065\AA \ absorption lines in near-peak spectra. These events also show an absorption line near $\lambda$6150\AA  \ whose nature is still debated \citep{2016ApJ...820...75P, 2016ApJ...827...90L}.
Normal SNe Ic are overall similar to SNe Ib but by definition they do not show the three strong characteristic He I features near peak.

However, there still exist some problems with these traditional and well-developed classification criteria. The first issue is that this classification is not quantitative, due to many historical reasons. 
The definition of Type I SNe is based on unquantified spectroscopic distinctions (i.e, whether silicon and helium absorption lines exist or not) around their peak luminosity. Though such an empirical criterion works well for most conditions, it is often subjective or controversial (see Section \ref{sec:subc} and \ref{sec:app}).	
Meanwhile, the present definition of SNe Ic is based on lack of prominent hydrogen and helium features. Such a negative definition leaves a question whether all objects that are not SNe Ia, Ib or II belongs to the Type Ic class. In order to distinguish SNe Ic from other SNe accurately, we have to give a positive definition of this class.

Furthermore, the nature of some Type I SN spectroscopic features is still not confirmed. In particular, the absorption near $\lambda$6150\AA \ in SN Ib spectra seems to be associated with H$\alpha$ \citep{2016ApJ...820...75P, 2016ApJ...827...90L}, while it may be due to Si II in SN Ic peak spectra. 

Therefore, our goal in this project is to find a quantitative classification criterion for Type I SNe, just like William Wilson Morgan, Philip C. Keenan and Edith Kellman have done in 1943 to introduce the Yerkes stellar spectral classification \citep{1943yerkes}.
One option is to use the line depth technique introduced by \citet{2012MNRAS.425.1819S}. By measuring line depths of indicative lines, we aim to classify these SNe with a quantitative criterion on their absolute or relative line intensity, with different types located at different positions on a diagram. 
While there are many absorption lines that could probably play the role of indicators in SN spectra, we find that the conspicuous absorption lines near $\lambda$6150\AA\ as well as O I $\lambda 7774$ \AA\ can be found in almost every peak spectrum of Type I SNe. A preliminary result is presented in \citet{2016arXiv161109353G} and shows these two absorption lines may be good indicators, however, the sample size in that work is fairly small.

Here we plan to present a robust result that Type I SNe can be reclassified using the line depths of OI $\lambda 7774$ \AA \ and the absorption at $\lambda 6150$ \AA \ in rest wavelength around their maximum light. The SN spectra sample we adopted consists of 146 SNe Ia (including 3 SNe 2002cx-like, 1 SN 1991T-like,  21 SNe 1991bg-like, 8 SNe 1999aa-like and a single other unclassified peculiar object, namely SN2000cx), 12 SNe Ib, 19 SNe Ic (including 5 SNe Ic-BL objects) and 4 SNe Ib/c spectra around their maximum light respectively. Compared to the preliminary result mentioned above, we have increased the sample size by an order of magnitude. All the SNe Ia we utilized in this paper are from the Berkeley Supernova Ia Program (BSNIP), while some of their corresponding maximum-light time or peak spectra are from other publications. All public SNe Ib and Ic peak spectra can be found on the Weizmann Interactive Supernova data REPository \citep[WISeREP,][]{Yaron2012}\footnote{\href{https://wiserep.weizmann.ac.il/}{https://wiserep.weizmann.ac.il/} }. We include data for several PTF objects kindly provided in advance of publication (Fremling et al., in preparation).

In Section \ref{sec:SNData} of this paper, we introduce the SN sample in detail and describe our selection criteria. The analysis methods we employed are presented in Section \ref{sec:ana}. Results are then presented and discussed in Section \ref{sec:res}.


\section{Supernovae Data}
\label{sec:SNData}
\subsection{SNe Ia}
\label{sec:SNeIa}
Though the total number of SN Ia discoveries is quite large, we only use BSNIP objects in this paper. The reasons we chose BSNIP are as follows.

First, BSNIP events are all located in low-redshift galaxies, which means at relatively low distances and thus probably have higher quality spectra. In addition, these SNe Ia are well-studied objects. Among the 582 SN Ia objects presented in \citet{2012MNRAS.425.1789S}, 199 have well-calibrated light curves, which can provide accurate phase information to 584 spectra. Since we want to study the spectroscopic features of Type I SNe at maximum light, it is crucial to know their dates of maximum. In addition, BSNIP spectra also have robust and uniform flux calibration and telluric correction. 


In this paper, the major source of object subclasses and maximum dates is \citet{2012MNRAS.425.1789S}. Each Type Ia SN subtype is directly adopted from the BSNIP I data release\footnote{\href{http://heracles.astro.berkeley.edu/sndb/info\#DownloadDatasets(BSNIP,LOSS)}{http://heracles.astro.berkeley.edu/sndb/info\#DownloadDatasets(BSNIP,LOSS)}}. 
132 of the total 146 SN Ia spectra we utilized are also from BSNIP I. Phase information for 584 spectra has been provided in that paper. We combined this set with other papers studying SN Ia light curves, for example, \citet{2012AJ....143..126B}, \citet{2011AJ....142..156S}, as well as \citet{2011AJ....142...74K} which studies one specific SN Ia (SN2001ay), and enlarge the sample size of BSNIP SN Ia peak dates. All the maximum dates are with respect to B-band maximum brightness in rest-frame days.

After assembling a large sample of BSNIP SNe Ia with maximum dates, we match them to the observation dates of spectra in BSNIP I. The quality cut criteria are as follows:

\indent 1. The rest wavelength range of each spectrum should cover the range from 5800\AA \ to 7800\AA, which contains both of the absorption regions we will measure.\\
\indent 2. The observation date of each spectrum should fall within 5 days prior to or later than B-band maximum. This is our definition of peak spectra in this paper.\\ 
\indent 3. If there are more than a single peak spectrum for a specific object, we will only use the one taken closest to its B-band maximum brightness. If the closest one is of fairly low Signal to Noise Ratio (SNR), we will turn to the second closest one and inspect its quality to decide whether we adopt it or not, and so on.\\  
\indent 4. If there are no corresponding peak spectra in the BSNIP I spectra package, we will turn to WISeREP to find whether there are other peak spectra in this database. The quality cut criteria are the same as 1 to 3 listed above.

After this step we find that 146 SNe Ia have corresponding peak spectra which can satisfy all the criteria listed above. Table \ref{tab:ia1} shows subtype, observation date and phase information of each SN Ia object we utilize in this paper.

Totally there are 6 subtypes for SNe Ia in Table \ref{tab:ia1}, including 114 normal Ia (referred to as ``SN Ia'' in the table; containing SN2005ao and SN2008s5, whose subtypes remain undetermined), 1 Ia-1991T, 19 Ia-1991bg, 8 Ia-1999aa, 3 Ia-2002cx and 1 Ia-pec, following the classifications by \citet{2012MNRAS.425.1789S}. SNe Ia-1999aa were classified as Ia-1991T objects in some previous publications \citep[e.g.][]{2001ApJ...546..734L} and these two subtypes show a great similarity in spectra except for Ca II H\&K absorption and Si II $\lambda$6355\AA\ line depth before maximum \citep{2004AJ....128..387G}. Due to this reclassification, the sample size of Ia-1991T objects has been sharply cut down. The Ia-pec subtype only contain SN2000cx, which is pointed out to be peculiar in its light curve features and spectral evolution properties \citep{2001PASP..113.1178L,Silverman2013}.

\begin{figure*}[ht]
  \centering
  \includegraphics[width=\linewidth]{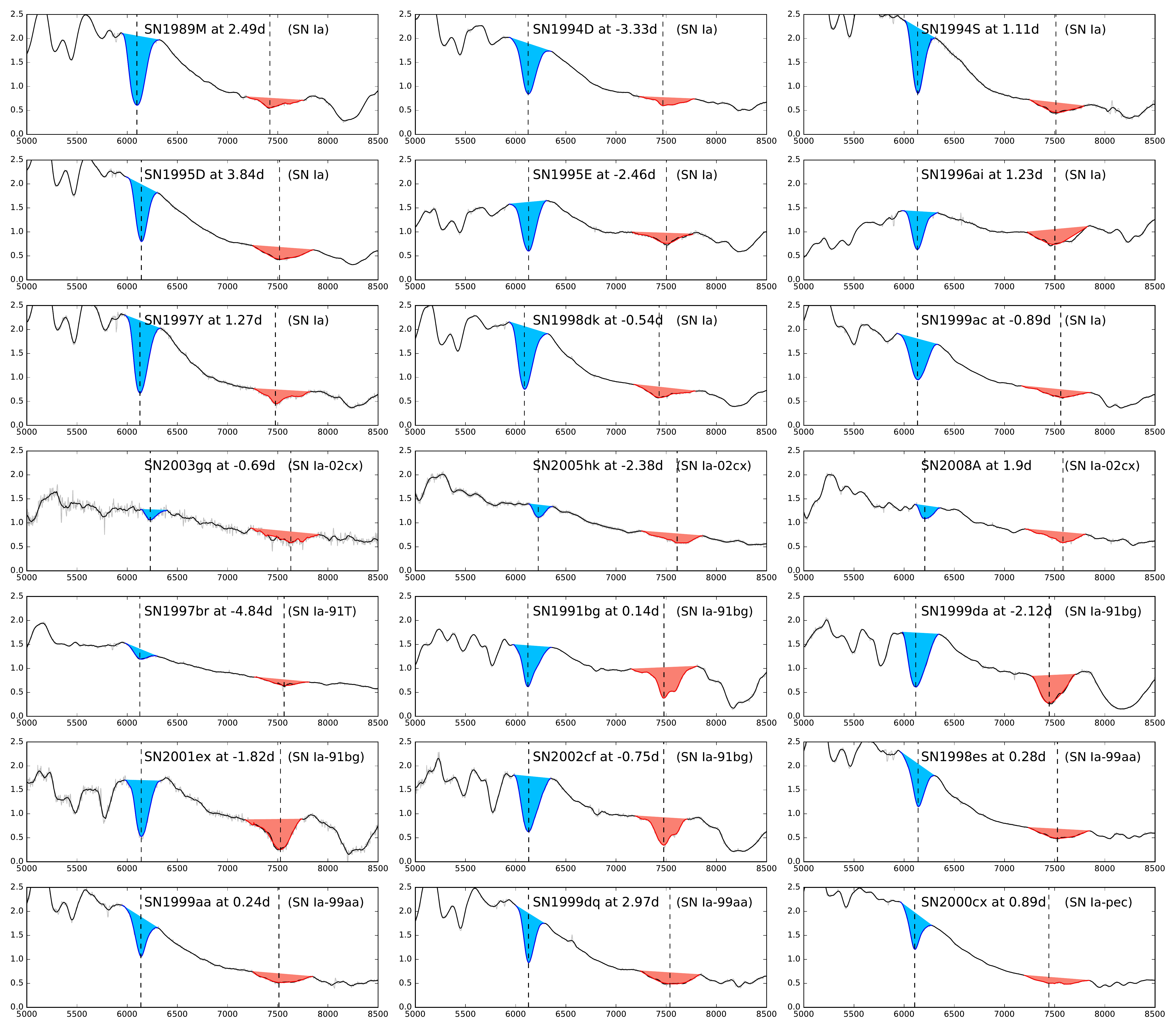}
  \caption{\footnotesize This figure shows 21 SN Ia spectra around their maximum light. In each subplot, the silver solid line (mostly covered by the black solid line) is the original spectrum (without smoothing, telluric line correcting and fitting), while the black solid line is the peak spectrum smoothed by a Savitzky-Golay filter. The blue and red solid lines represent the $\lambda$6150 \AA \ and O I absorption line regions, respectively. 
  Both these regions have been de-blended from telluric lines and other absorption lines in the SN spectrum and fitted by a polynomial with a degree of 9 (if the curve is not smooth enough). The light blue and red shading show the difference between the pseudo-continuum and the absorption lines at the $\lambda$6150 \AA \ and O I regions, respectively, while black vertical dashed lines represent the locations of maximum line depth (see its definition at Section \ref{sec:method}). In all of these spectra, wavelengths are calibrated to the rest frame according to their host galaxy heliocentric redshifts. The flux density is normalized to an arbitrary scale. As is shown in the figure, all of these peak spectra show a strong absorption at 6150 \AA \ , which is widely believed to be due to Si II absorption lines in SN Ia ejecta. Name, subtype and phase information of each spectrum could be found on each subplot. \label{fig:iaspec}} 
\end{figure*}

Maximum date references and spectra references for these 146 Type Ia SNe are also presented in Table \ref{tab:ia1}. The heliocentric redshifts of these events are directly acquired from WISeREP and \citet{2012MNRAS.425.1789S}, which are originally from the NASA/IPAC Extragalactic Database (NED\footnote{\href{https://ned.ipac.caltech.edu/}{https://ned.ipac.caltech.edu/}}). In order to show the average quality of  all 146 SN Ia spectra we analyze, we present 21 of them, including 9 SNe Ia (normal), 3 SNe Ia-2002cx, 1 SN Ia-1991T, 4 SNe Ia-1991bg, 3 SNe Ia-1999aa and 1 SN Ia-pec in Figure \ref{fig:iaspec}. Other than the original spectra (silver line in the plot), we also plot the smooth spectra (black line) using a Savitzky-Golay filter and mark the two absorption regions we will measure and discuss in later sections of this paper.

\begin{deluxetable*}{cccccccc}
\tablecaption{Summary of 146 SNe Ia Spectra\label{tab:ia1}}
\tabletypesize{\footnotesize}
\tablewidth{\linewidth}
\tablehead{\colhead{SN Name} & \colhead{Type} & \colhead{Redshift} & \colhead{Obseration Date (MJD)} & \colhead{Phase\tnm{a}(day)} & \colhead{Maximum Reference\tnm{b}} & \colhead{Spectra Reference\tnm{c}} & \colhead{Instrument and Telescope\tnm{d}} }
\startdata
SN1989M & SN Ia & 0.005 & 47716.0 & 2.49 & (1) & (1) & (1) \\
SN1994D & SN Ia & 0.001 & 49429.0 & -3.33 & (1) & (1) & (2) \\
SN1994S & SN Ia & 0.015 & 49519.34 & 1.11 & (1) & (1) & (2) \\
SN1995D & SN Ia & 0.007 & 49772.43 & 3.84 & (1) & (1) & (2) \\
SN1995E & SN Ia & 0.012 & 49772.31 & -2.46 & (1) & (1) & (2) \\
SN1996ai & SN Ia & 0.003 & 50256.33 & 1.23 & (2) & (1) & (2) \\
SN1997Y & SN Ia & 0.016 & 50488.35 & 1.27 & (1) & (1) & (2) \\
SN1998dk & SN Ia & 0.013 & 51056.31 & -0.54 & (1) & (1) & (2) \\
SN1999ac & SN Ia & 0.009 & 51249.47 & -0.89 & (1) & (1) & (2) \\
SN1999cp & SN Ia & 0.009 & 51368.2 & 4.91 & (1) & (1) & (2) \\
SN1999gd & SN Ia & 0.018 & 51517.58 & -1.12 & (1) & (1) & (3) \\
SN1999gh & SN Ia & 0.008 & 51517.61 & 4.12 & (1) & (1) & (3) \\
SN2000cp & SN Ia & 0.034 & 51722.32 & 2.92 & (1) & (1) & (2) \\
SN2000cw & SN Ia & 0.03 & 51753.36 & 4.81 & (1) & (1) & (2) \\
SN2000dg & SN Ia & 0.038 & 51793.44 & 4.66 & (1) & (1) & (2) \\
SN2000dk & SN Ia & 0.018 & 51813.48 & 1.0 & (1) & (1) & (2) \\
SN2000dm & SN Ia & 0.015 & 51813.24 & -1.63 & (1) & (1) & (2) \\
SN2000dn & SN Ia & 0.032 & 51823.33 & -0.94 & (1) & (1) & (2) \\
SN2001ay & SN Ia & 0.03 & 52025.91 & 3.91 & (4) & (5) & (9) \\
SN2001az & SN Ia & 0.041 & 52029.0 & -3.24 & (1) & (1) & (2) \\
SN2001ba & SN Ia & 0.029 & 52029.0 & -4.64 & (1) & (1) & (2) \\
SN2001bf & SN Ia & 0.016 & 52045.44 & 1.22 & (1) & (1) & (2) \\
SN2001bg & SN Ia & 0.007 & 52039.72 & -0.28 & (2) & (6) & (8) \\
SN2001bp & SN Ia & 0.095 & 52045.46 & 0.51 & (1) & (1) & (2) \\
SN2001br & SN Ia & 0.021 & 52054.26 & 3.48 & (1) & (1) & (2) \\
SN2001cp & SN Ia & 0.022 & 52089.42 & 1.39 & (1) & (1) & (2) \\
SN2001da & SN Ia & 0.017 & 52106.46 & -1.12 & (1) & (1) & (2) \\
SN2001ep & SN Ia & 0.013 & 52202.47 & 2.83 & (1) & (1) & (2) \\
SN2001fe & SN Ia & 0.013 & 52228.44 & -0.99 & (1) & (1) & (2) \\
SN2002aw & SN Ia & 0.026 & 52326.56 & 2.1 & (1) & (1) & (2) \\
SN2002bf & SN Ia & 0.024 & 52340.4 & 2.97 & (1) & (1) & (3) \\
SN2002bo & SN Ia & 0.004 & 52355.17 & -1.08 & (1) & (1) & (2) \\
SN2002bz & SN Ia & 0.037 & 52372.48 & 4.92 & (1) & (1) & (2) \\
SN2002cd & SN Ia & 0.01 & 52384.51 & 1.1 & (1) & (1) & (2) \\
SN2002ck & SN Ia & 0.03 & 52401.45 & 3.64 & (1) & (1) & (2) \\
SN2002de & SN Ia & 0.028 & 52433.37 & -0.32 & (1) & (1) & (2) \\
SN2002eb & SN Ia & 0.028 & 52495.32 & 1.68 & (1) & (1) & (2) \\
SN2002ef & SN Ia & 0.024 & 52495.51 & 4.7 & (1) & (1) & (2) \\
SN2002er & SN Ia & 0.009 & 52520.22 & -4.58 & (1) & (1) & (2) \\
SN2002eu & SN Ia & 0.038 & 52520.44 & -0.06 & (1) & (1) & (2) \\
SN2002ha & SN Ia & 0.014 & 52580.14 & -0.85 & (1) & (1) & (2) \\
SN2002he & SN Ia & 0.025 & 52586.0 & 0.29 & (1) & (1) & (3) \\
SN2003U & SN Ia & 0.028 & 52674.55 & -2.55 & (1) & (1) & (2) \\
SN2003cq & SN Ia & 0.033 & 52737.45 & -0.15 & (1) & (1) & (2) \\
SN2003gn & SN Ia & 0.035 & 52856.27 & 3.26 & (1) & (1) & (2) \\
SN2003he & SN Ia & 0.025 & 52879.37 & 2.71 & (1) & (1) & (2) \\
SN2003hv & SN Ia & 0.006 & 52892.0 & 2.9 & (2) & (2) & (6) \\
SN2003iv & SN Ia & 0.034 & 52935.47 & 1.76 & (1) & (1) & (2) \\
SN2004as & SN Ia & 0.031 & 53080.61 & -4.36 & (1) & (1) & (3) \\
SN2004bl & SN Ia & 0.017 & 53137.19 & 4.61 & (1) & (1) & (2) \\
SN2004dt & SN Ia & 0.02 & 53241.4 & 1.38 & (1) & (1) & (2) \\
SN2004fu & SN Ia & 0.009 & 53328.12 & 2.43 & (1) & (1) & (2) \\
SN2004gs & SN Ia & 0.027 & 53356.45 & 0.44 & (1) & (1) & (2) \\
SN2004gu & SN Ia & 0.046 & 53358.0 & -4.65 & (1) & (1) & (2) \\
SN2005M & SN Ia & 0.022 & 53409.0 & -1.41 & (1) & (1) & (2) \\
SN2005W & SN Ia & 0.009 & 53412.2 & 0.59 & (1) & (1) & (4) \\
SN2005ag & SN Ia & 0.079 & 53413.6 & 0.52 & (1) & (1) & (3) \\
SN2005am & SN Ia & 0.008 & 53440.3 & 4.47 & (1) & (1) & (3) \\
SN2005ao & SN Ia & 0.038 & 53442.54 & 0.52 & (1) & (1) & (2) \\
SN2005bc & SN Ia & 0.012 & 53471.47 & 1.55 & (1) & (1) & (2) \\
SN2005bo & SN Ia & 0.014 & 53478.23 & -0.07 & (2) & (3) & (5) \\
SN2005cf & SN Ia & 0.006 & 53532.21 & -1.19 & (1) & (1) & (2) \\
SN2005de & SN Ia & 0.015 & 53597.34 & -0.75 & (1) & (1) & (2) \\
SN2005di & SN Ia & 0.025 & 53608.41 & 0.49 & (1) & (1) & (2) \\
SN2005dv & SN Ia & 0.01 & 53624.17 & -0.57 & (1) & (1) & (2) \\
SN2005el & SN Ia & 0.015 & 53647.53 & 1.22 & (1) & (1) & (2) \\
SN2005ki & SN Ia & 0.02 & 53706.65 & 1.62 & (1) & (1) & (4) \\
SN2005lz & SN Ia & 0.04 & 53736.38 & 0.58 & (1) & (1) & (4) \\
SN2005ms & SN Ia & 0.025 & 53741.4 & -1.88 & (1) & (1) & (2) \\
SN2005na & SN Ia & 0.026 & 53740.33 & 0.03 & (1) & (1) & (2) \\
SN2006D & SN Ia & 0.009 & 53759.43 & 3.7 & (1) & (1) & (2) \\
SN2006N & SN Ia & 0.014 & 53759.29 & -0.9 & (1) & (1) & (2) \\
SN2006X & SN Ia & 0.005 & 53788.41 & 3.15 & (1) & (1) & (2) \\
SN2006ax & SN Ia & 0.017 & 53827.14 & 0.34 & (2) & (3) & (5) \\
SN2006br & SN Ia & 0.025 & 53851.21 & 1.69 & (1) & (3) & (5) \\
SN2006bt & SN Ia & 0.032 & 53860.44 & 2.27 & (1) & (1) & (2) \\
SN2006bu & SN Ia & 0.081 & 53860.32 & 4.22 & (1) & (1) & (2) \\
SN2006cj & SN Ia & 0.068 & 53883.35 & 3.43 & (1) & (1) & (2) \\
SN2006cm & SN Ia & 0.016 & 53883.42 & -1.15 & (1) & (1) & (2) \\
SN2006cq & SN Ia & 0.048 & 53891.31 & 2.0 & (1) & (1) & (2) \\
SN2006ef & SN Ia & 0.018 & 53971.45 & 3.2 & (1) & (1) & (2) \\
SN2006ej & SN Ia & 0.02 & 53971.43 & -3.7 & (1) & (1) & (2) \\
SN2006et & SN Ia & 0.022 & 53997.35 & 3.29 & (1) & (1) & (2) \\
SN2006gj & SN Ia & 0.028 & 54003.48 & 4.7 & (1) & (1) & (2) \\
SN2006kf & SN Ia & 0.021 & 54038.39 & -3.05 & (1) & (1) & (2) \\
SN2006or & SN Ia & 0.021 & 54062.65 & -2.79 & (1) & (1) & (3) \\
SN2006ot & SN Ia & 0.053 & 54062.36 & -1.34 & (3) & (1) & (3) \\
SN2006sr & SN Ia & 0.024 & 54094.24 & 2.69 & (1) & (1) & (3) \\
SN2007A & SN Ia & 0.017 & 54113.21 & 2.37 & (1) & (1) & (3) \\
SN2007F & SN Ia & 0.024 & 54126.53 & 3.23 & (1) & (1) & (2) \\
SN2007O & SN Ia & 0.036 & 54122.66 & -0.33 & (1) & (1) & (4) \\
SN2007S & SN Ia & 0.014 & 54144.24 & 0.14 & (2) & (3) & (5) \\
SN2007af & SN Ia & 0.005 & 54172.54 & -1.25 & (1) & (1) & (2) \\
SN2007bc & SN Ia & 0.021 & 54200.32 & 0.61 & (1) & (1) & (2) \\
SN2007bd & SN Ia & 0.031 & 54207.09 & 0.59 & (2) & (3) & (5) \\
SN2007bm & SN Ia & 0.006 & 54228.2 & 3.7 & (2) & (3) & (6) \\
SN2007bz & SN Ia & 0.022 & 54216.42 & 1.65 & (1) & (1) & (2) \\
SN2007ca & SN Ia & 0.014 & 54228.25 & 1.55 & (2) & (3) & (6) \\
SN2007ci & SN Ia & 0.018 & 54244.26 & -1.71 & (1) & (1) & (2) \\
SN2007co & SN Ia & 0.027 & 54265.47 & 0.85 & (1) & (1) & (2) \\
SN2007fb & SN Ia & 0.018 & 54288.49 & 1.95 & (1) & (1) & (2) \\
SN2007fr & SN Ia & 0.051 & 54301.29 & -1.25 & (1) & (1) & (2) \\
SN2007gi & SN Ia & 0.005 & 54326.17 & -0.35 & (1) & (1) & (2) \\
SN2007gk & SN Ia & 0.027 & 54326.24 & -1.72 & (1) & (1) & (2) \\
SN2007hj & SN Ia & 0.014 & 54348.25 & -1.23 & (1) & (1) & (2) \\
SN2007on & SN Ia & 0.006 & 54416.47 & -3.0 & (1) & (1) & (3) \\
SN2007s1 & SN Ia & 0.069 & 54406.13 & -1.23 & (1) & (1) & (2) \\
SN2008ar & SN Ia & 0.026 & 54537.0 & 2.83 & (1) & (1) & (2) \\
SN2008bf & SN Ia & 0.025 & 54556.17 & 1.57 & (2) & (3) & (5) \\
SN2008cl & SN Ia & 0.063 & 54603.31 & 4.24 & (1) & (1) & (2) \\
SN2008ec & SN Ia & 0.016 & 54673.36 & -0.24 & (1) & (1) & (2) \\
SN2008ei & SN Ia & 0.038 & 54673.31 & 3.29 & (1) & (1) & (2) \\
SN2008s1 & SN Ia & 0.028 & 54612.0 & 0.49 & (1) & (1) & (2) \\
SN2008s5 & SN Ia & 0.069 & 54731.3 & 1.26 & (1) & (1) & (2) \\
SN2003gq & SN Ia-02cx & 0.021 & 52847.29 & -0.69 & (1) & (1) & (2) \\
SN2005hk & SN Ia-02cx & 0.013 & 53682.19 & -2.38 & (1) & (1) & (4) \\
SN2008A & SN Ia-02cx & 0.016 & 54480.3 & 2.2 & (2) & (4)& (2) \\
SN1997br & SN Ia-91T & 0.007 & 50554.43 & -4.84 & (1) & (1) & (2) \\
SN1991bg & SN Ia-91bg & 0.003 & 48603.0 & 0.14 & (1) & (1) & (1) \\
SN1999da & SN Ia-91bg & 0.012 & 51368.36 & -2.12 & (1) & (1) & (2) \\
SN2001ex & SN Ia-91bg & 0.026 & 52202.43 & -1.82 & (1) & (1) & (2) \\
SN2002cf & SN Ia-91bg & 0.015 & 52384.33 & -0.75 & (1) & (1) & (2) \\
SN2002dk & SN Ia-91bg & 0.018 & 52442.35 & -1.23 & (1) & (1) & (2) \\
SN2002fb & SN Ia-91bg & 0.016 & 52530.29 & 0.98 & (1) & (1) & (2) \\
SN2003Y & SN Ia-91bg & 0.017 & 52674.23 & -1.74 & (1) & (1) & (2) \\
SN2005er & SN Ia-91bg & 0.026 & 53648.23 & -0.26 & (1) & (1) & (4) \\
SN2006bz & SN Ia-91bg & 0.027 & 53860.22 & -2.44 & (1) & (1) & (2) \\
SN2006cs & SN Ia-91bg & 0.023 & 53891.33 & 2.28 & (1) & (1) & (2) \\
SN2006em & SN Ia-91bg & 0.019 & 53980.44 & 4.16 & (1) & (1) & (2) \\
SN2006gt & SN Ia-91bg & 0.045 & 54003.35 & 3.08 & (1) & (1) & (2) \\
SN2006ke & SN Ia-91bg & 0.017 & 54032.4 & 2.36 & (1) & (1) & (2) \\
SN2007N & SN Ia-91bg & 0.013 & 54122.58 & 0.44 & (1) & (1) & (4) \\
SN2007al & SN Ia-91bg & 0.012 & 54172.29 & 3.39 & (1) & (1) & (2) \\
SN2007ax & SN Ia-91bg & 0.007 & 54185.06 & -2.04 & (3) & (3) & (7) \\
SN2007ba & SN Ia-91bg & 0.038 & 54200.47 & 2.14 & (1) & (1) & (2) \\
SN2008bt & SN Ia-91bg & 0.015 & 54572.0 & -1.08 & (1) & (1) & (2) \\
SN2008dx & SN Ia-91bg & 0.023 & 54646.22 & 2.46 & (1) & (1) & (2) \\
SN1998es & SN Ia-99aa & 0.011 & 51142.25 & 0.28 & (1) & (1) & (2) \\
SN1999aa & SN Ia-99aa & 0.015 & 51232.24 & 0.24 & (1) & (1) & (2) \\
SN1999dq & SN Ia-99aa & 0.014 & 51438.52 & 2.97 & (1) & (1) & (2) \\
SN2001eh & SN Ia-99aa & 0.037 & 52172.48 & 3.26 & (1) & (1) & (2) \\
SN2005eq & SN Ia-99aa & 0.029 & 53654.38 & 0.66 & (1) & (1) & (2) \\
SN2006S & SN Ia-99aa & 0.032 & 53772.52 & 2.99 & (1) & (1) & (2) \\
SN2006cz & SN Ia-99aa & 0.042 & 53906.27 & 1.12 & (1) & (1) & (2) \\
SN2008Z & SN Ia-99aa & 0.021 & 54512.46 & -2.29 & (1) & (1) & (2) \\
SN2000cx & SN Ia-pec & 0.008 & 51753.4 & 0.89 & (1) & (1) & (2) 
\enddata
\tablenotetext{a}{We adopt phase information from \citet{2012MNRAS.425.1789S}, or derive the phase from the corresponding maximum date references.}
\tablenotetext{b}{Maximum date reference:  (1) \citet{2012MNRAS.425.1789S}; (2) \citet{2012AJ....143..126B}; (3) \citet{2011AJ....142..156S}; (4) \citet{2011AJ....142...74K}.}
\tablenotetext{c}{Spectra reference: (1) \citet{2012MNRAS.425.1789S}; (2) \citet{Led09}; (3) \citet{2013ApJ...773...53F}; (4) \citet{2014ApJ...786..134M}; (5) \citet{2011AJ....142...74K}; (6) Obtained from Avishay Gal-Yam \citep[used for SN2001bg's classification in ][]{2001IAUC.7622....2G}} 
\tablenotetext{d}{Instrument and Telescope: (1) UV Schmidt (Shane 3 m); (2) KAST (Shane 3 m); (3) LRIS (Keck 10 m); (4) DEIMOS (Keck 10 m); (5) WFCCD (Ir\'en\'ee du Pont 2.5 m); (6) IMACS (Magellan 6.5 m); (7) Boller \& Chivens Spectrograph (Ir\'en\'ee du Pont 2.5 m); (8) FOSC (Wise 1m); (9) MMT-Blue (MMT 6.5m).} 
\end{deluxetable*}

\subsection{SNe Ib and Ic} 
\label{sec:ibc}
For SNe Ib and Ic, including regular SNe Ib, regular SNe Ic, peculiar SNe Ic-BL and transitional or unclear type Ib/c SNe, we use a similar method to select appropriate spectra from WISeREP to study. Since the total number of SN Ib and Ic discoveries is far less than SNe Ia \citep[e.g.][]{GalYam2013PASP}, we decide to use all the data fulfilling our quality cut criteria we could find in WISeREP.

We extracted a list of SN maximum dates from WISeREP, which also contain their classification information. Then we validate the list and enlarge the sample size using additional publications and data (see following subsections for details). After this step we match the peak dates with the observation dates of the corresponding spectra in WISeREP. We find that totally 12 regular SNe Ib, 14 regular SNe Ic, 5 SNe Ic-BL and 4 transitional or unclear type Ib/c SNe have qualified peak spectra for analyzing.

\subsubsection{Dates of Maximum}
Initially the maximum date data are directly obtained from WISeREP. These peak dates should be validated before being utilized.

The band of maximum date and its reference are vacant in this database, making peak date information in WISeREP currently unreliable.
In addition, the sample size is a bit low for a statistical study. Quite a large fraction of SNe Ib and Ic spectra in WISeREP are from the CfA Supernova Data Archive. Though many of these objects are calibrated with multi-color light curves (for example, in \citet{Modjaz2014AJ}, 44 of 73 SNe have well-determined dates of maximum light to determine the phase of each spectrum), many of these spectra were taken with the FAST Instrument on the FLWO 1.5m, SAO, whose wavelength range poorly covers the O I $\lambda$7774\AA\ absorption we would like to measure. Therefore, before we continue, we need to improve this list of maximum dates.

In order to do this necessary improvement, we use maximum light data from a number of publications, which could be found in Table \ref{tab:ibc1} for each SN.
We also add 5 unpublished PTF Stripped-envelope Core-Collapse Supernovae (CCSNe) into this table. 3 of these maximum dates are generated from the PTF photometry, including 2 SNe Ib (PTF10fbv and PTF10feq) and 1 unclear SN Ib/c (PTF12hvv).

For most of these SNe, we adopt a maximum date based on V-band or g-band light curves (mainly for PTF and SDSS SNe). The reason why we adopted V-band maximum instead of B-band is that B-band maximum dates are more likely to be vacant than V-band in our major references (for example, \citet{Bianco2014ApJS} present 
 SN2005az V-band maximum dates but without B-band ones; \citet{Drout2011ApJ} only present V-band and R-band maximum dates). Since the time interval between B-band and V-band maximum is typically about -2 days \citep[from data presented in][]{Bianco2014ApJS}, which is much smaller than the 10 days time range of ``peak spectra" we defined, it makes little difference. 
For the sample whose maximum dates are cited from \citet{Lyman2016MNRAS}, we adopt their bolometric peak dates. The detailed band information for each SN peak date can be found in Table \ref{tab:ibc1}.

\subsubsection{Subclassification}
\label{sec:subc}
The initial subclassification information for our SN Ib and Ic sample are also from WISeREP. Here we are going to discuss this matter in detail.

In this paper we mainly care about quantitative classification criteria of SNe Ia, Ib and Ic. Because the total discovery rate of SNe Ib and Ic is quite small (compared to SNe Ia), we could not split the sample to many subclasses of SNe Ib or Ic. Therefore, we focus on regular SNe Ib, SNe Ic and broad-line SNe Ic-BL.

Historically, when astronomers determined the type of a newly discovered SN Ib or Ic, sometimes there was ambiguity and uncertainty in the classification. For example, SN2005az, which we claim is a SN Ic event \citep{2005IAUC.8504....3Q}, used to be classified as a SN Ib near peak brightness according to its spectrum \citep{2005ATel..451....1A}. This kind of confusion has not been rare at all since Type Ic SNe were distinguished from Type Ib. Therefore, we have to inspect the classification of each object we study in this paper. The detailed reference for each SN classification could be found in Table \ref{tab:ibc1}. 

In addition to the regular Type Ib, Ic and Ic-BL objects in the table, we also add 4 so-called SNe Ib/c to the list. Three SNe Ib/c (SN1999ex, SN2009jf, SN2013ge) are either of a transitional type between Ib and Ic or those whose classification remains controversial.
SN1999ex was characterized by the lack of hydrogen lines, weak optical He I lines, and strong He I $\lambda\lambda$10830\AA\ and 20581\AA, thus providing an example of an intermediate case between pure Ib and Ic supernovae \citep{2002AJ....124..417H}. 
SN2009jf was classified as a young SN Ib by \citet{2009CBET.1955....1K} and \citet{2009CBET.1955....2S}, based on early spectra obtained on September 29, 2009, while its spectroscopic similarity with the He-poor Type Ic SN 2007gr \citep{2011MNRAS.416.3138V} motivates us to consider it as a SN Ib/c event \citep{2016arXiv161109353G}.
SN2013ge is classified as a SN Ib/c due to the lack of a strong Si II $\lambda$6355\AA\ feature, while the detection of weak helium features indicates that it could generally be classified as Type Ic \citep{Drout2016}.

In Section \ref{sec:app}, we will try to use our quantitative classification criteria to clearly classify these events. 

\begin{deluxetable*}{ccccccc}

\tabletypesize{\footnotesize}
\tablewidth{\linewidth}
\tablecaption{Summary of SNe Ib and Ic's classification and maximum dates\label{tab:ibc1}}
\tablehead{\colhead{SN Name} & \colhead{Type} & \colhead{Redshift} & \colhead{Peak Date (MJD)} & \colhead{Band\tnm{a}} & \colhead{Classification Reference\tnm{b}} & \colhead{Maximum Reference\tnm{c}} }
\startdata
PTF10fbv & SN Ib & 0.056 & 55291 & r & ... & ... \\
PTF10feq & SN Ib & 0.0279 & 55291 & r & ... & ... \\
PTF10qif & SN Ib & 0.064 & 55406.37 & g & (1) & (1) \\
SN1990I & SN Ib & 0.0097 & 48010 & -- & (18) & (10) \\
SN1999dn & SN Ib & 0.0094 & 51419.3 & V & (10) & (11) \\
SN2004dk & SN Ib & 0.0052 & 53238 & V & (23) & (12) \\
SN2005bf & SN Ib & 0.0189 & 53497.73 & V & (25) & (13) \\
SN2007Y & SN Ib & 0.0047 & 54165.6 & V & (2) & (2) \\
SN2007nc & SN Ib & 0.0868 & 54393.21 & g & (26) & (3) \\
SN2007uy & SN Ib & 0.007 & 54481.34 & V & (27) & (13) \\
SN2008D & SN Ib & 0.0065 & 54494.24 & V & (28) & (14) \\
iPTF13bvn & SN Ib & 0.0045 & 56475.24 & B & (31) & (15) \\
PTF12hvv & SN Ib/c & 0.029 & 56165 & r & ... & ... \\
SN1999ex & SN Ib/c & 0.0114 & 51501.2 & V & (11) & (16) \\
SN2009jf & SN Ib/c & 0.008 & 55122.32 & V & (12) for Ib; (13) for Ic & (17) \\
SN2013ge & SN Ib/c & 0.0044 & 56618.6 & V & (7) & (7) \\
PTF11rka & SN Ic & 0.0744 & 55922.47 & g & (1) & (1) \\
SN1990B & SN Ic & 0.0075 & 47909 & -- & (6) & (6) \\
SN1991N & SN Ic & 0.0033 & 48348 & -- & (19) & (18) \\
SN1994I & SN Ic & 0.0015 & 49451.4 & V & (20) & (19),(20) \\
SN2004aw & SN Ic & 0.0163 & 53090.95 & V & (21) & (22) \\
SN2004dn & SN Ic & 0.0126 & 53230.5 & V & (22) & (12) \\
SN2005az & SN Ic & 0.0085 & 53473.36 & V & (24) & (13) \\
SN2006aj & SN Ic & 0.033 & 53794.23 & V & (15) & (13) \\
SN2007gr & SN Ic & 0.0017 & 54338.5 & V & (16) & (23) \\
SN2007qx & SN Ic & 0.0804 & 54419.57 & g & (3) & (3) \\
SN2010bh & SN Ic & 0.0593 & 55279.3 & Bolometric & (29) & (24) \\
SN2011bm & SN Ic & 0.022 & 55677.2 & Bolometric & (30) & (24) \\
PTF12gzk & SN Ic & 0.0138 & 56148.24 & V & (8) & (8) \\
iPTF15dtg & SN Ic & 0.054 & 57353.5 & g & (5) & (5) \\
PTF10bzf & SN Ic-BL & 0.0498 & 55257 & R & (17) & (25) \\
PTF11img & SN Ic-BL & 0.158 & 55777 & r & ... & ... \\
SN1998bw & SN Ic-BL & 0.0085 & 50945.2 & V & (4) & (4) \\
SN2002ap & SN Ic-BL & 0.0021 & 52313.42 & V & (14),(32),(33) & (21) \\
SN2012ap & SN Ic-BL & 0.0122 & 55976.4 & V & (9) & (9) \\
\enddata
\tablenotetext{a}{The vacancy (--) in the Band column means that the band used for determining peak dates is not mentioned in the corresponding reference (for SN1990B, SN1990I, SN1991N).}
\tablenotetext{b}{Classification Reference: (1) \citet{Prentice2016}; (2) \citet{2009ApJ...696..713S}; (3) \citet{Taddia2015}; (4) \citet{1998Natur.395..670G}; (5) \citet{Taddia2016}; (6) \citet{1993ApJ...409..162V}; (7) \citet{Drout2016}; (8) \citet{2012ApJ...760L..33B}; (9) \citet{2015ApJ...799...51M}; (10) \citet{2000ApJ...540..452D}; (11) \citet{2002AJ....124..417H}; (12) \citet{2011MNRAS.413.2583S}; (13) \citet{2011MNRAS.416.3138V}; (14) \citet{2002MNRAS.332L..73G}; (15) \citet{2006ApJ...645L..21M}; (16) \citet{2008ApJ...672L..99C}; (17) \citet{2011ApJ...741...76C}; (18) \citet{1990IAUC.5032....2P}; (19) \citet{1991IAUC.5234....1F}; (20) \citet{1994IAUC.5981....2K}; (21) \citet{2004IAUC.8331....2F}; (22) \citet{2004IAUC.8381....1G}; (23) \citet{2004IAUC.8404....1F}; (24) \citet{2005IAUC.8504....3Q}; (25) \citet{2005IAUC.8522....2M}; 
(26) \citet{2007CBET.1104....1B}; (27) \citet{2008CBET.1191....2B}; (28) \citet{2008CBET.1222....1M}; (29) \citet{2010CBET.2228....1C}; (30) \citet{2011CBET.2695....1R}; (31) \citet{2013ATel.5142....1M}; (32) \citet{2002ApJ...572L..61M}; (33) \citet{2003PASP..115.1220F}}

\tablenotetext{c}{Maximum Reference: (1) \citet{Prentice2016}; (2) \citet{2009ApJ...696..713S}; (3) \citet{Taddia2015}; (4) \citet{1998Natur.395..670G}; (5) \citet{Taddia2016}; (6) \citet{1993ApJ...409..162V}; (7) \citet{Drout2016}; (8) \citet{2012ApJ...760L..33B}; (9) \citet{2015ApJ...799...51M}; (10) \citet{2004AandA...426..963E}; (11) \citet{2011MNRAS.411.2726B}; (12) \citet{Drout2011ApJ}; (13) \citet{Bianco2014ApJS}; (14) \citet{2009ApJ...702..226M}; (15) \citet{2014MNRAS.445.1932S}; (16) \citet{2002AJ....124.2100S}; (17) \citet{2011MNRAS.416.3138V}; (18) \citet{1999AandAS..139..531B}; (19) \citet{1996AJ....111..327R}; (20) \citet{2001AJ....121.1648M}; (21) \citet{2003PASP..115.1220F}; (22) \citet{2006MNRAS.371.1459T}; (23) \citet{2009AandA...508..371H}; (24) \citet{Lyman2016MNRAS}; (25) \citet{2011ApJ...741...76C}}
\end{deluxetable*}

\subsubsection{Quality Cuts and Spectra}
After constructing a list of SNe Ib and Ic maximum dates, we match them with observation dates of spectra in WISeREP. 

The quality cut criteria are similar to those we used to select SN Ia spectra (see Section \ref{sec:SNeIa}). We require a rest-frame wavelength range that extends from 5800\AA\ to 7800\AA\ (except for a single spectrum of SN2009jf that nevertheless covers the O I $\lambda$7774\AA\ region), and spectroscopic observation obtained within 5 days around maximum dates. 
This is followed by quality inspection. If the closest spectrum to maximum light is of fairly low quality, for example, low SNR or bad host-galaxy subtraction, we will turn to the second closest one within 5 days to maximum date, and so on.

\begin{figure*}[ht]
  \centering
  \includegraphics[width=\linewidth]{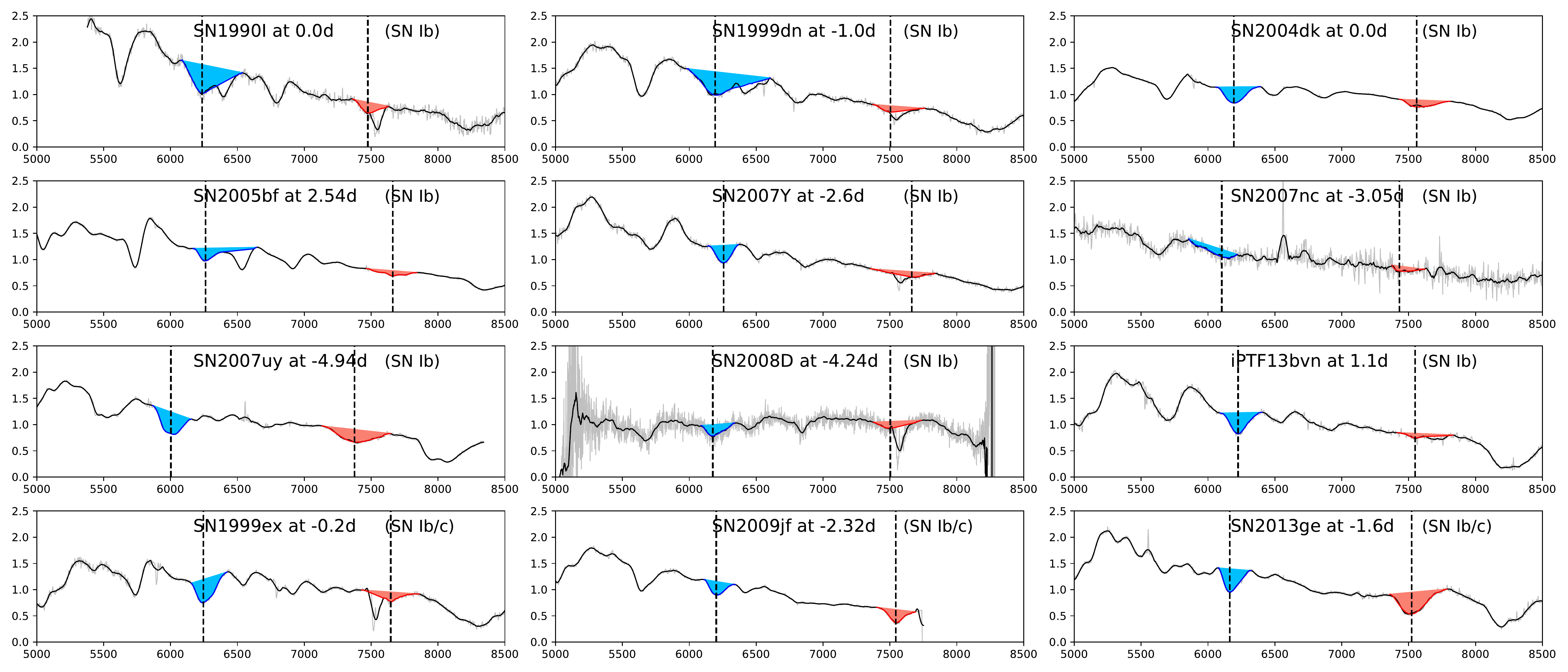}
  \caption{\footnotesize This figure shows all of the 9 published SN Ib and 3 SN Ib/c spectra around their maximum light. In each subplot, the silver, black, red, and blue solid line, as well as the colored filled region and black vertical dashed lines are the same as described in Figure \ref{fig:iaspec}. It is clear that compared with the SN Ia spectra in Figure \ref{fig:iaspec}, SN Ib peak spectra show relatively shallow $\lambda$6150 \AA \ and O I absorption. The $\lambda$6150 \AA \  line is often blended with the He I line at $\lambda$6678 \AA \ , which sometimes complicates our measurement of the $\lambda$6150 \AA \ line depth. Name, type and phase information of each spectrum could be found on each subplot. \label{fig:ibspec}} 
\end{figure*}

\begin{figure*}[ht]
  \centering
  \includegraphics[width=\linewidth]{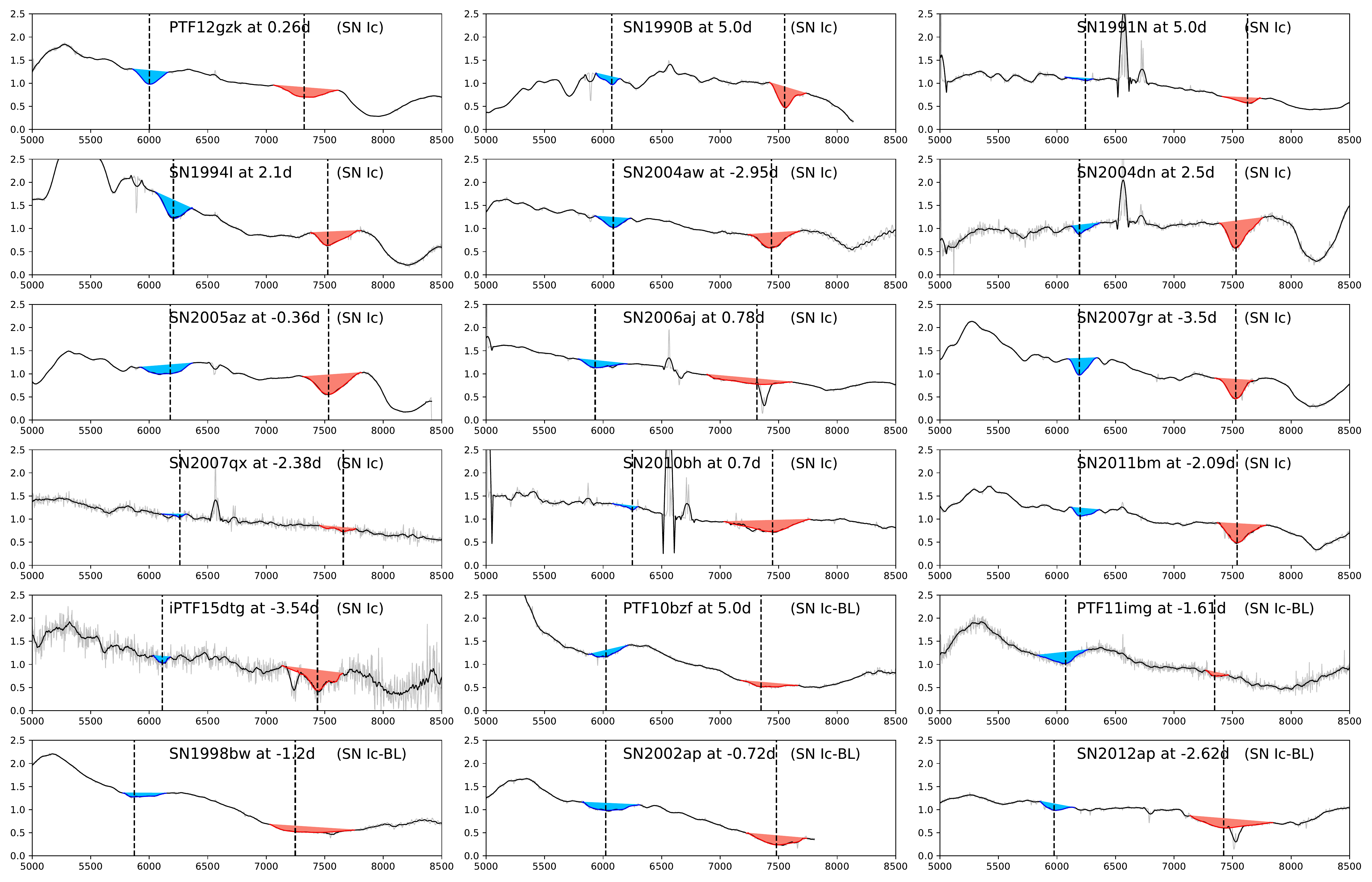}
  \caption{\footnotesize This figure shows all of the 13 published SN Ic and 5 SN Ic-BL spectra around their maximum light. In each subplot, the silver, black, red, and blue solid line, as well as the shading and black vertical dashed lines are the same as described in Figure \ref{fig:iaspec}. Note that settings the SN1998bw and PTF10bzf blue endpoints of the $\lambda$6150 \AA \ line is tricky, making the analysis of these two object less reliable. Therefore we only report them for completeness.
  Compared with SN Ib spectra in Figure \ref{fig:ibspec}, we can find that SNe Ic have a relatively strong O I absorption, while their $\lambda$6150 \AA \  line depths are still much shallower than those of SN Ia presented in Figure \ref{fig:iaspec}. Name, type and phase information of each spectrum could be found on each subplot.\label{fig:icspec}} 
\end{figure*}

Table \ref{tab:ibcspec} shows the summary of all the SN Ib and Ic spectra we adopt, including observation dates, phases, instruments and telescopes, and corresponding references. In order to demonstrate the quality of SN Ib and Ic spectra we analyze, we present 30 spectra in Figure \ref{fig:ibspec} (SNe Ib) and \ref{fig:icspec} (SNe Ic). Other than the original spectra (silver line in the plot), we also plot the smooth spectra (black line) using a Savitzky-Golay filter and mark the two absorption regions we will measure and discuss in the Section \ref{sec:ana} of this paper.

\begin{deluxetable*}{cccccc}
\tablecaption{Summary of SNe Ib and Ic's spectra\label{tab:ibcspec}}
\tabletypesize{\footnotesize}
\tablewidth{\textwidth}
\tablehead{\colhead{SN Name} & \colhead{Type} & \colhead{Obseration Date (MJD)} & \colhead{Phase (day)} & \colhead{Instrument and Telescope\tnm{a}} & \colhead{Spectra Reference\tnm{b}} }
\startdata
PTF10fbv & SN Ib & 55291 & 0 & (1) & PTF \\
PTF10feq & SN Ib & 55291 & 0 & (1) & PTF \\
PTF10qif & SN Ib & 55409 & 2.63 & (3) & PTF\\
SN1990I & SN Ib & 48010 & 0 & (5) & (1) \\
SN1999dn & SN Ib & 51418.3 & -1.0 & (6) & (2) \\
SN2004dk & SN Ib & 53238 & 0 & (2) & (21) \\
SN2005bf & SN Ib & 53501 & 2.54 & (1) & (3) \\
SN2007Y & SN Ib & 54163 & -2.6 & (8) & (4) \\
SN2007nc & SN Ib & 54390.16 & -3.05 & (8) & (5) \\
SN2007uy & SN Ib & 54476.4 & -4.94 & (9) & (6) \\
SN2008D & SN Ib & 54490 & -4.24 & (3) & (7) \\
iPTF13bvn & SN Ib & 56476.34 & 1.1 & (10) & (20) \\
PTF12hvv & SN Ib/c & 56160 & -5 & (3) & PTF \\
SN1999ex & SN Ib/c & 51501 & -0.2 & (8) & (8) \\
SN2009jf & SN Ib/c & 55120.0 & -2.32 & (12) & (9) \\
SN2013ge & SN Ib/c & 56617.0 & -1.6 & (13) & (10) \\
PTF11rka & SN Ic & 55921.65 & -0.82 & (1) & PTF\\
SN1990B & SN Ic & 47914 & 5 & (14) & (11) \\
SN1991N & SN Ic & 48353 & 5 & (14) & (11) \\
SN1994I & SN Ic & 49453.5 & 2.1 & (9) & (6) \\
SN2004aw & SN Ic & 53088 & -2.95 & (15) & (12) \\
SN2004dn & SN Ic & 53233 & 2.5 & (14) & (21) \\
SN2005az & SN Ic & 53473 & -0.36 & (7) & (21) \\
SN2006aj & SN Ic & 53795.01 & 0.78 & (16) & (13) \\
SN2007gr & SN Ic & 54335 & -3.5 & -- & (14) \\
SN2007qx & SN Ic & 54417.19 & -2.38 & (8) & (5) \\
SN2010bh & SN Ic & 55280.0 & 0.7 & (17) & (15) \\
SN2011bm & SN Ic & 55675.11 & -2.09 & (18) & (16) \\
PTF12gzk & SN Ic & 56148.5 & 0.26 & (14) & (19) \\
iPTF15dtg & SN Ic & 57349.96 & -3.54 & (15) & (21) \\
PTF10bzf & SN Ic-BL & 55262 & 5 & (1) & (22) \\
PTF11img & SN Ic-BL & 55775.39 & -1.61 & (1) & (23) \\
SN1998bw & SN Ic-BL & 50944.0 & -1.2 & (6) & Asiago SN Group \\
SN2002ap & SN Ic-BL & 52312.7 & -0.72 & (4) & (17) \\
SN2012ap & SN Ic-BL & 55973.78 & -2.62 & (11) & (18) \\
\enddata
\tablenotetext{a}{Instrument and Telescpoe: (1) LRIS (Keck 10m); (2) DBSP (P200); (3) ISIS (WHT 4.2m); (4) FOSC (Wise 1m); (5) EFOSC (ESO 2.2m); (6) DFOSC (Danish 1.54m); (7) LRS (HET 10m); (8) EMMI (NTT 3.58m); (9) MMT-Blue (MMT 6.5m); (10) FLOYDS (FTN 2m); (11) RSS (SALT); (12) AFOSC (Ekar 1.82m); (13) Hectospec (MMT 6.5m); (14) KAST (Shane 3m); (15) DOLORES (TNG 3.58m); (16) FORS2 (VLT-UT1 8.2m); (17) X-Shooter (VLT-UT2 8.2m); (18) ALFOSC (NOT 2.5m). The vacancy (--) in this column means that instrument and telescope information is not mentioned in either spectra reference paper or the WISeREP database.}
\tablenotetext{b}{Spectra Reference: (1) \citet{2004AandA...426..963E}; (2) \citet{2011MNRAS.411.2726B}; (3) \citet{2006ApJ...641.1039F}; (4) \citet{2009ApJ...696..713S}; (5) \citet{2011AandA...526A..28O}; (6) \citet{Modjaz2014AJ}; (7) \citet{2009ApJ...692L..84M}; (8) \citet{2002AJ....124..417H}; (9) \citet{2011MNRAS.416.3138V}; (10) \citet{Drout2016}; (11) \citet{2001AJ....121.1648M}; (12) \citet{2006MNRAS.371.1459T}; (13) \citet{2006Natur.442.1011P}; (14) \citet{Valenti2008ApJ}; (15) \citet{2012ApJ...753...67B}; (16) \citet{2012ApJ...749L..28V}; (17) \citet{2002MNRAS.332L..73G}; (18) \citet{2015ApJ...799...51M}; (19) \citet{2012ApJ...760L..33B} ; (20) \citet{Frem16}; (21) Gal-Yam et al. (in preparation); (21) \citet{Taddia2016}; (22) \citet{2011ApJ...741...76C}; (23) \citet{Corsi2016ApJ}. Here ``PTF'' indicates unpublished PTF data. The spectrum of SN1998bw is acquired from Asiago SN group.}

\end{deluxetable*}

\section{Analysis}
\label{sec:ana}
\subsection{Line Depth Measurement Method}
\label{sec:method}
In this section we describe our spectroscopic analysis method. Since we want to measure the line depth of the absorption regions around $\lambda$6150\AA\ and $\lambda$7500\AA\ relative to the pseudo-continuum, we adopt the line-depth technique presented in \citet{2012MNRAS.425.1819S} which was initially developed to measure the absorption region features of the BSNIP SN Ia sample. We will describe the technical details, including our improvement of their method in the following paragraphs. We then show a comparison between the \citet{2012MNRAS.425.1819S} and our measurement on BSNIP SNe Ia Si II and O I absorption line depths.

\subsubsection{Initial Processing}
For each of the spectra we listed in Table \ref{tab:ia1} and Table \ref{tab:ibcspec}, we correct the wavelength to rest-frame using the host-galaxy redshifts presented in Table \ref{tab:ia1} and Table \ref{tab:ibc1}. Since interstellar reddening does not affect absorption line depth, we did not correct this effect. 

Due to the high expansion velocity of the ejecta, the typical spectrum of a SN should not contain narrow absorption features. Therefore, it is reasonable to smooth the spectra and only retain broad spectral features for analysis. Each spectrum is smoothed using a Savitzky-Golay smoothing filter \citep{sgfilter}. The window size (frame length) and polynomial order for each spectrum is decided by its average wavelength interval, which is:
\begin{equation}
w = 2\times \textrm{int}(\frac{50}{{\rm\Delta}\lambda})+1
\end{equation}
\begin{equation}
o =  \textrm{max}(3,\frac{w-1}{2})
\end{equation}
where $w$ is the window size, $o$ is the polynomial order and ${\rm\Delta}\lambda$ is the average wavelength interval (in \AA). By smoothing the spectra, we can effectively remove small scale abnormal flux values which could be due to cosmic rays, galactic or other possible contamination. The filtering process could help us deal with low SNR spectra rather than drop them. 

\subsubsection{Pseudo-Continuum and Line Depth}
One of the most difficult steps in analyzing SN spectra is determining pseudo-continua for the absorption lines. Though the local minimum for the absorption is easy to determine both by eye and by machine, it is difficult to define the endpoints of an absorption line due to several reasons. 
As a result of the ejecta high expansion velocity, the absorption lines in SN spectra are quite broad, and absorptions due to different transitions can blend together (for example, Si II $\lambda$6355\AA\ may blend with He I $\lambda$6678\AA, which is common in SN Ib spectra). Furthermore, especially for SNe Ic-BL, the flux blueward of the Si II $\lambda$6355\AA\ absorption often does not reach a peak before blending with the absorption around $\lambda$5000\AA\ (for example, see SN1998ew's spectrum in Figure \ref{fig:icspec}). A similiar problem sometimes affectes O I $\lambda$7774\AA\ because the continuum could have a sharp slope. All these effects make absorption line endpoints hard to determine.

In order to determine the range of the absorption regions around $\lambda$6150\AA\ and O I $\lambda$7774\AA, we decided to determine the endpoints manually. Though this method could introduce human error and some subjective bias in the endpoint definition, if the measurements are conducted by only one person during a short period, the human error will mainly result in a systematic shift and the fluctuation between measurements of different spectra will be relatively small. Besides, we can also examine our measurements by comparing our line depth results with \citet{2012MNRAS.425.1819S}. We will present the result of this comparison at the end of this section.

\begin{figure*}[t]
 \centering
  \includegraphics[width=0.32\linewidth]{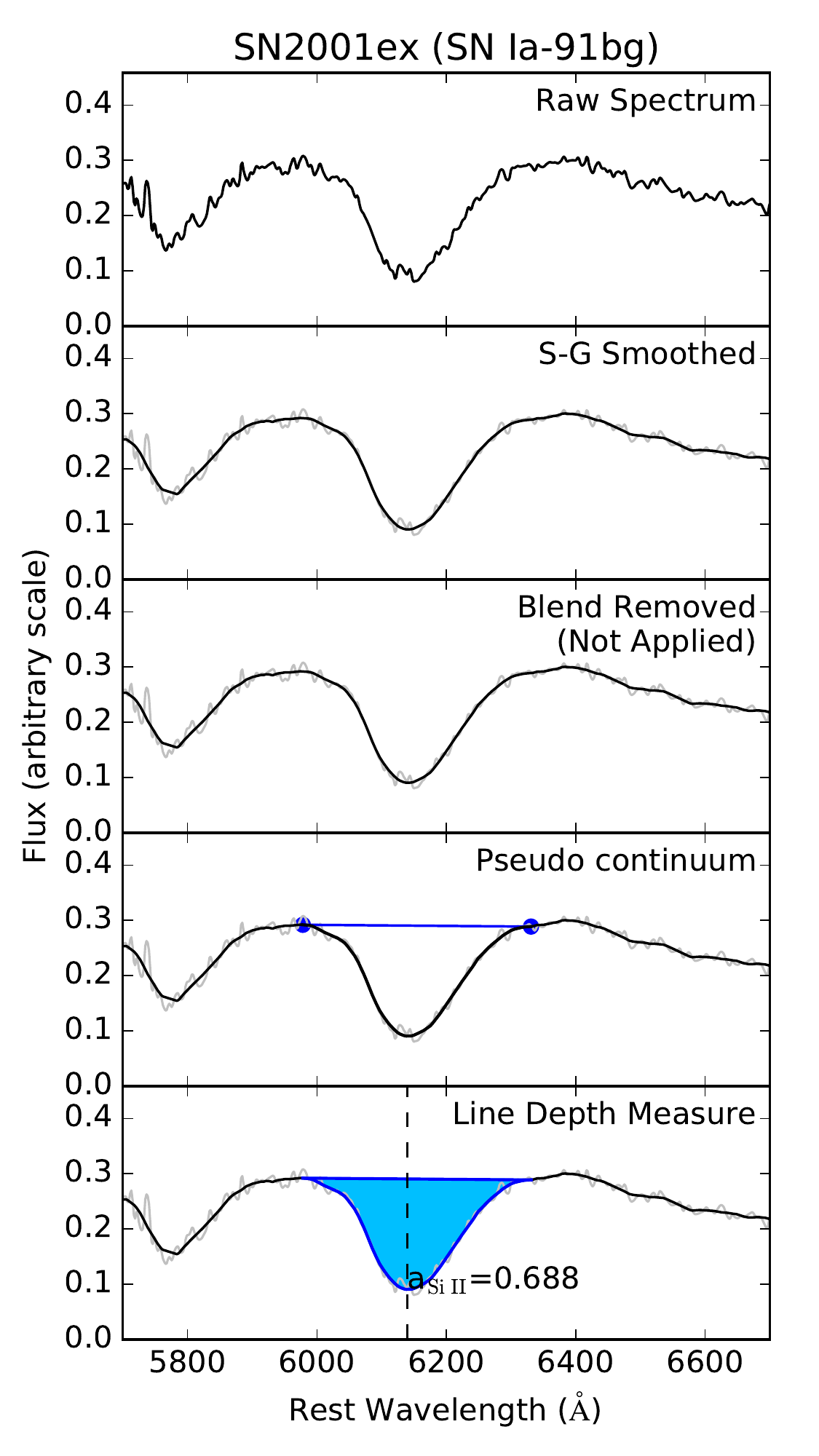}
  \includegraphics[width=0.32\linewidth]{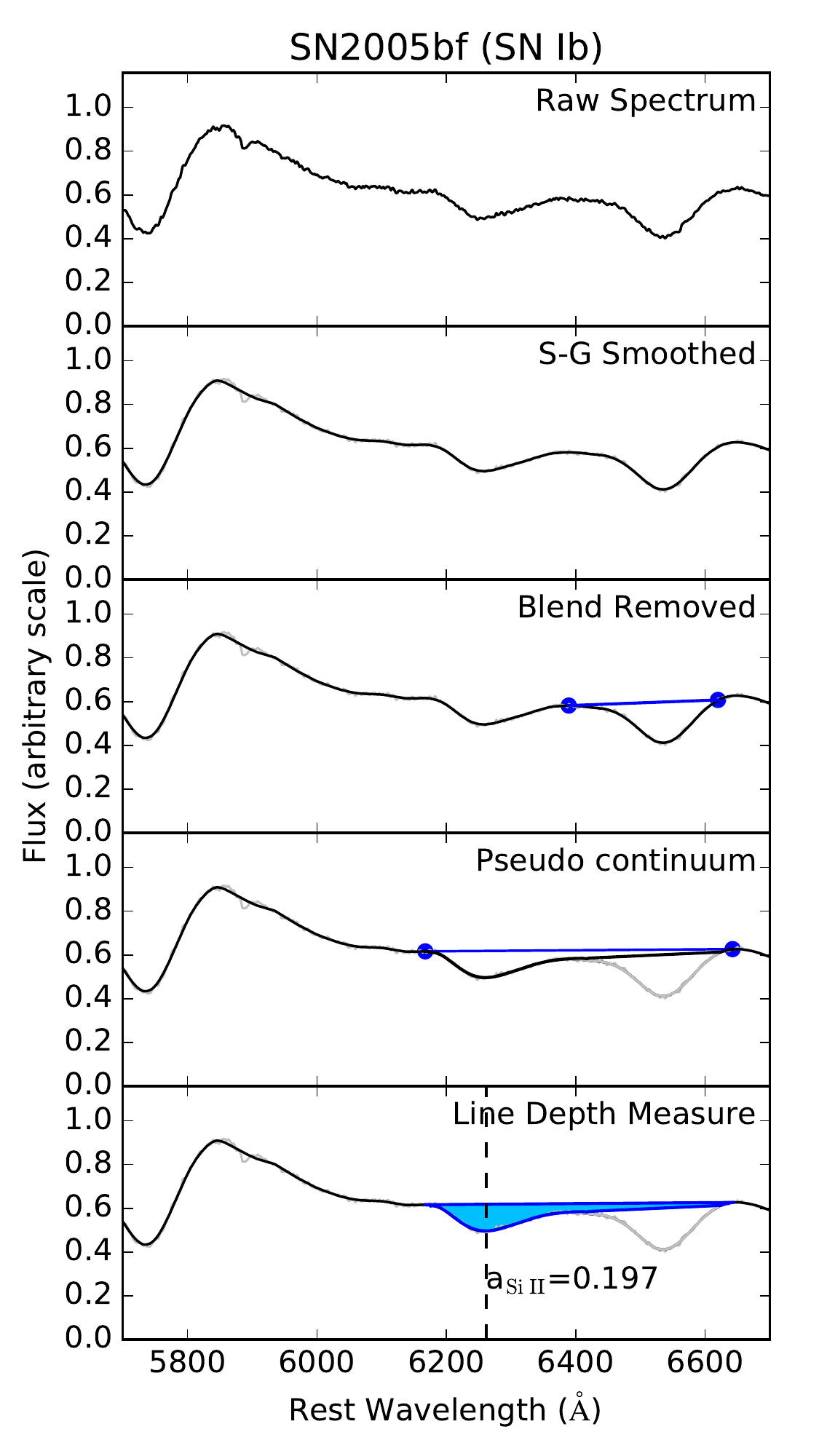}
  \includegraphics[width=0.32\linewidth]{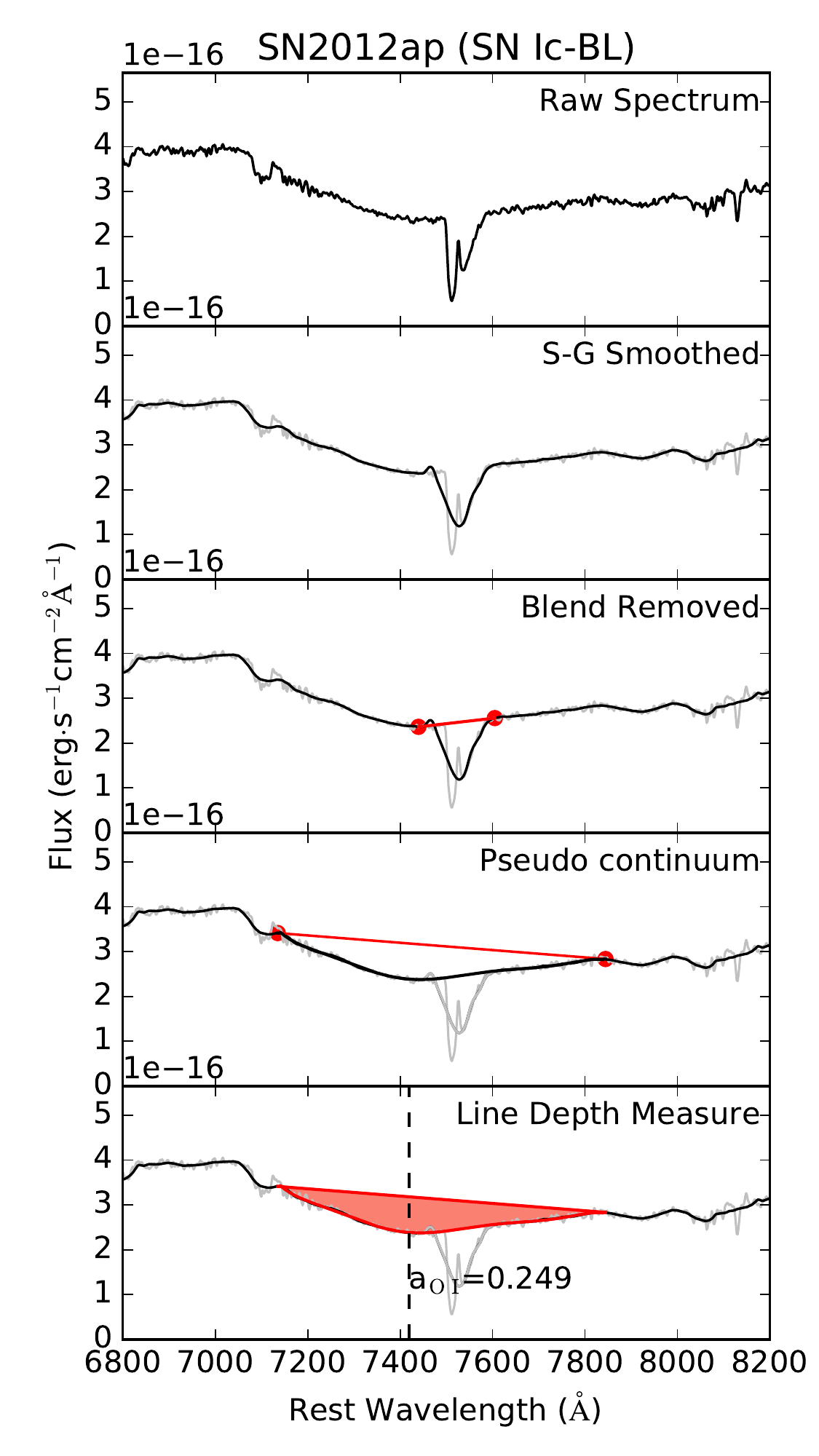}
  \caption{\footnotesize This figure shows the regular procedure of measuring an absorption line depth, including raw spectra, Savitzky-Golay smoothing, blend removal, pseudo-continuum determination and line depth measure (from top to bottom). Three example presented here are the SN2001ex Si II $\lambda$6355\AA\ absorption, the SN2005bf absorption around $\lambda$6200\AA\ (blended with He I $\lambda$6678\AA; this absorption may not be due to Si II) and the SN2012ap O I $\lambda$7774\AA\ absorption (heavily affected by the Fraunhofer A band around $\lambda$7594\AA ). Here ``raw spectrum" means the original spectrum we adopted to analyze, which has been well-reduced. In order to show every change we made to the raw spectra, we plot every previous step result as silver line on the background.\label{fig:process}}
\end{figure*}

In this work, the determination of the two absorption line endpoints of 181 Type I SNe are done by Fengwu Sun individually. For most cases, these endpoints present the closest peak (both blueward and redward) to the local minimum for the corresponding spectral feature, while for the situation that a nearby peak is quite far from the local minimum, we will choose the point where the spectral slope gets flat. If the absorption line is heavily blended with other lines, including other transitions in the SN photosphere, galactic emission lines, or even telluric lines, we will firstly remove this contaminating line with a linear sub-pseudo-continuum (simply replace the spectral feature with a line segment whose endpoints are the start and end points of this feature; Figure \ref{fig:process}) and then determine the endpoints by the method we mentioned above. Under this de-blending condition, the endpoints for the absorption line may not be the intrinsic ones any longer (e.g. SN2005bf in Figure \ref{fig:process}). Since we only care about the depth rather than the width, this de-blending trick will not bring negative effects to our measurements. In this way, line depth measurements are superior to (less sensitive than) pseudo-equivalent widths (pEWs).

Once the absorption line endpoints are determined, the pseudo-continua are calculated as a linear function with the two endpoints. If the spectrum is still not smooth within the corresponding wavelength range, we will fit this part of the spectrum using a polynomial model with a degree of 9. This step can enhance the precision of line depth measurements when the bottom of an absorption line is slightly fluctuating.

After finishing all the steps we mentioned above, We calculate the line depth of each absorption. The line depth ($a$) is defined as:

\begin{equation}
a = {\rm max}\big(1-\frac{F_\lambda}{F_{\lambda,c}}\big)
\end{equation}

Where $F_{\lambda,c}$ is the flux value of the pseudo-continuum, and $F_\lambda$ is the flux value of the spectrum after modification. Both $F_{\lambda,c}$ and $F_\lambda$ are functions of wavelength, and the line depth, i.e. the maximum for $1-{F_\lambda}/{F_{\lambda,c}}$ could always be found in a given wavelength range. We present three detailed examples of the line depth measurement procedure in Figure \ref{fig:process}.

\begin{table*}[!htp]
\caption{Summary of abnormal discrepancies in BSNIP SN Ia Line Depth Measurements \label{tab:abnormal}}
\centering
{
\footnotesize
\begin{tabular}{cccc||cccc}
\hline\hline
SN Name & Absorption & Line Depth in & Line Depth & SN Name & Absorption & Line Depth in & Line Depth \\
 & Line & \citet{2012MNRAS.425.1819S} & in this paper &  & Line & \citet{2012MNRAS.425.1819S} & in this paper \\
\hline\hline 
SN2000cp  & O I &  0.749  &  0.287 &  SN2001br  & Si II & 0.784  &  0.457 \\
SN2001ex  & O I &  0.963  &  0.711 &  SN2001ex  & Si II & 0.813  &  0.688 \\
SN2008s1  & O I &  0.598  &  0.406 &  SN2002aw  & Si II & 0.655  &  0.452 \\
SN2000dn  & Si II & 0.762  &  0.659 &  SN2002de  & Si II & 0.779  &  0.530 \\
\hline\hline
\end{tabular}
}
\end{table*}

\begin{figure*}[!htp]
 \centering
  \includegraphics[width=\linewidth]{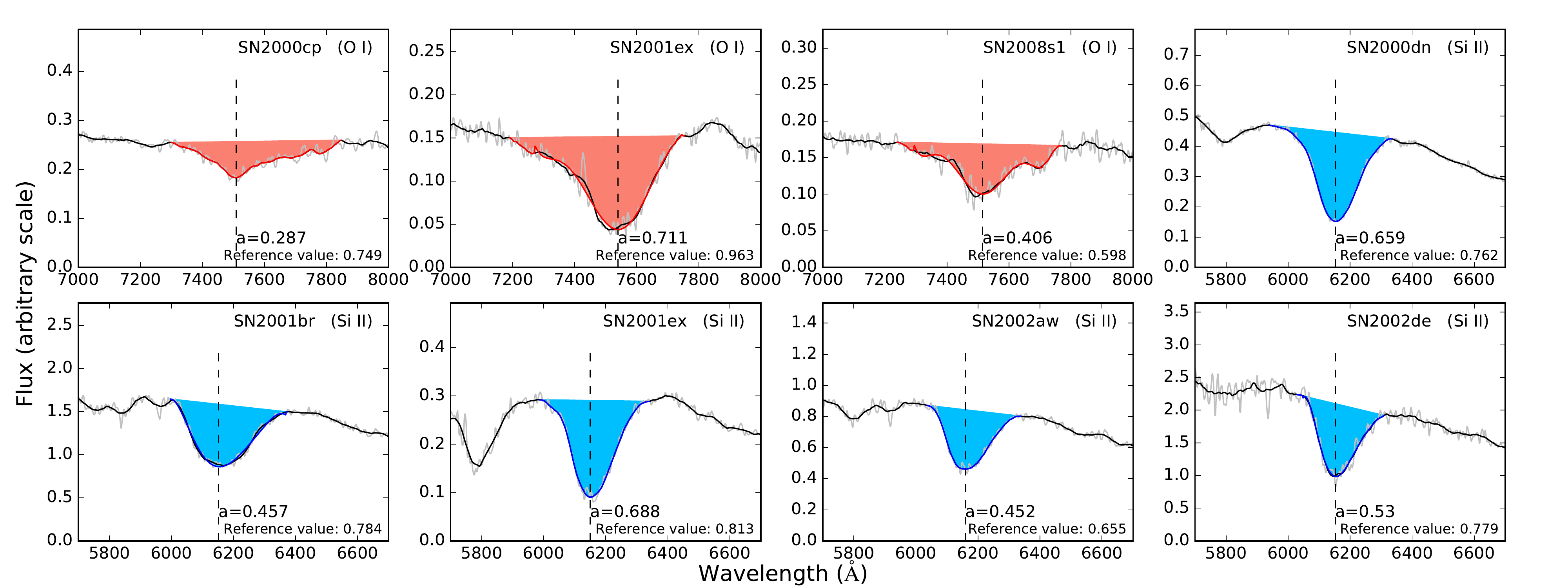}
  \caption{\footnotesize This figure shows all the 8 absorption lines whose depths are abnormally discrepant from the reference values. The silver, black, red, and blue solid line, as well as the colored filled region and black vertical dashed line are the same as described in Figure \ref{fig:iaspec}. Spectra are shifted such that the location of the Si II $\lambda$6355\AA\ maximum line depth is at 6150 \AA. Object name, line depth measurement result and reference value for each spectra are annotated on each subplot. By presenting this plot, we intend to demonstrate that our manual measurements are reasonable and convincing, in the cases that they differ from \citet{2012MNRAS.425.1819S} results. \label{fig6}} 
\end{figure*}

\begin{figure}[!h]
 \centering
  \includegraphics[width=1.0\linewidth]{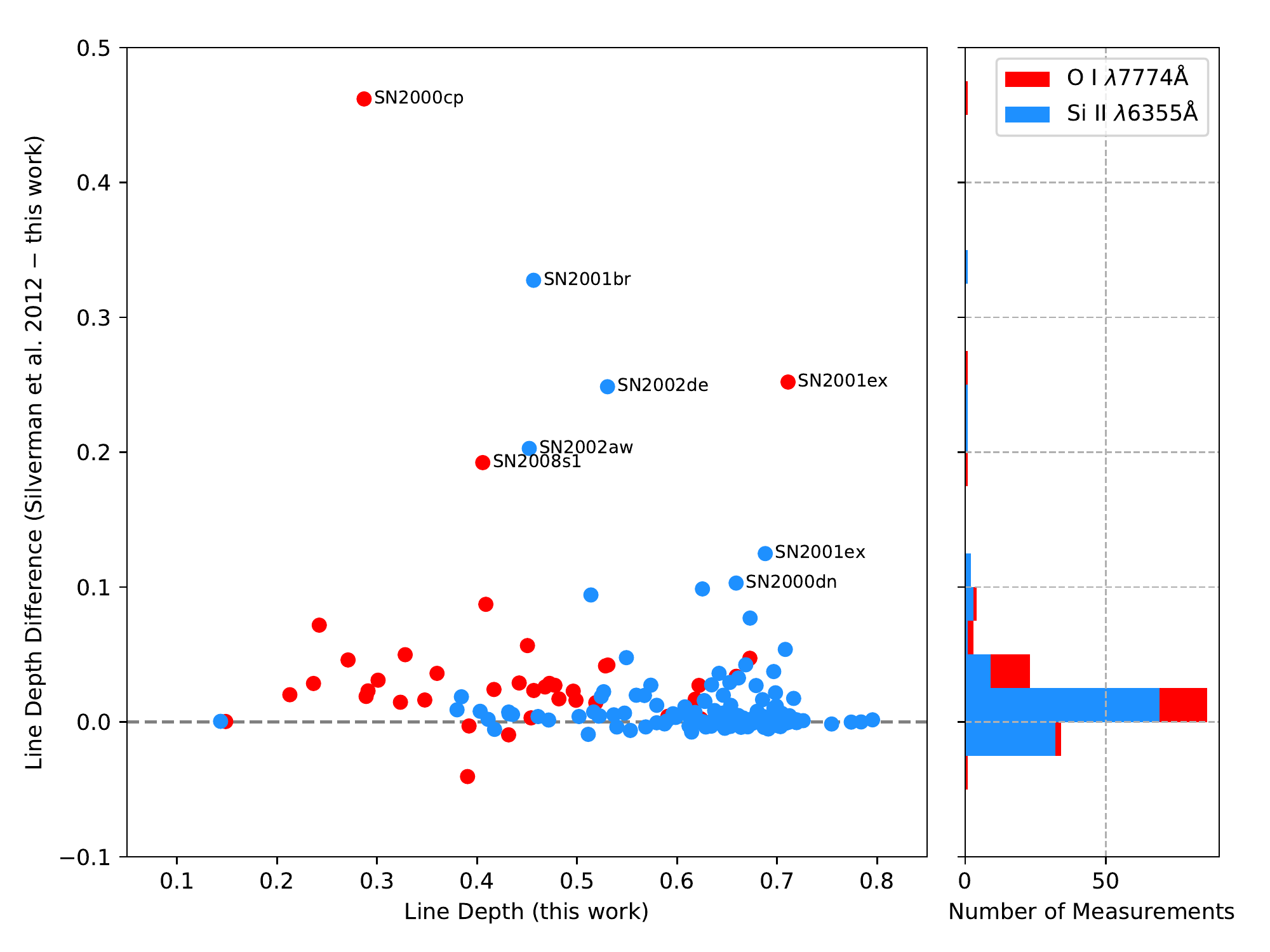}
  \caption{\footnotesize \textit{Left}: Scatter plot of line depth measurement differences between this paper and \citet{2012MNRAS.425.1819S}, where blue and red dots represent Si II $\lambda$6355\AA\ and O I $\lambda$7774\AA\ absorption line measurements respectively. For those objects whose differences are greater than 0.1, we also annotate their name on the plot. \quad \textit{Right}: A histogram of line depth measurement differences between the two works. The size of the bins along the vertical axis is 0.025. Data in each bin are stacked together.  \label{fig5}} 
\end{figure}

\subsection{Examination}
\label{sec:exa}
In order to assess the accuracy of our line depth result, we compare our measurements of BSNIP SNe Ia with \citet{2012MNRAS.425.1819S}. 119 Si II $\lambda$6355\AA\ and 40 O I $\lambda$7774\AA\ line depth measurements among all our 146 SN Ia spectra have counterparts in \citet{2012MNRAS.425.1819S}. Therefore, we plot a scatter diagram to observe the consistency between the two measurements (Figure \ref{fig5}, left panel). We also calculate the difference between the two independent measurements of each absorption line. A histogram of line depth measurement differences for Si II $\lambda$6355\AA\ and O I $\lambda$7774\AA\ is shown in the right panel of Figure \ref{fig5}.

Our interactive line depth measurement results are overall lower than \citet{2012MNRAS.425.1819S}, however the mean value of the differences ($-0.024$) is small ($\sim$4\% of the average line depth). Among all these 159 measurements, 144 (90.6\%) are within 0.05 and 120 (75.5\%) are within 0.025 of the \citet{2012MNRAS.425.1819S} results. Only 8 (5.0\%) of these discrepancies are larger than 0.1 and these values are large enough to attract our attention. In Table \ref{tab:abnormal} we list all these 8 abnormal measurement discrepancies. We also plot their spectra to demonstrate the validity of our measurements (see Figure \ref{fig6}). Inspection does not show any reason to suspect our result.

Alternatively, performing blind repeat measurements on both copies of the same spectra and other peak spectra of the same object taken a few days apart result in consistent measurements ($<10\%$ difference in values), which would not influence any of the results shown in the following section.


\section{Result and Discussion}
\label{sec:res}
In this section we intend to present the line depth measurement results and analyze their statistical properties. Since a preliminary result of Type I SN quantitative classification criteria has been formulated in \citet{2016arXiv161109353G}, here we will show an upgraded result with a larger sample size and more detailed subclass information. 

\subsection{Line Depth Measurement Results and Statistical Analysis}
In Table \ref{tab:measure} we present part of the line depth measurement results of 32 among all 181 Type I SN spectra, including 6 normal SNe Ia, 3 Ia-2002cx, 1 Ia-1991T, 4 Ia-1991bg, 4 Ia-1999aa, 1 Ia-pec, 3 regular SNe Ib, 4 SNe Ib/c, 4 regular SNe Ic and 2 Ic-BL (the full set of measurement results is available online). In addition to the results of the $\lambda6150$\AA\ and O I $\lambda$7774\AA\  line depths, we also calculate their ratio which would be used for analysis in the following paragraphs.

\begin{table*}
\caption{Summary of  Line Depth Measurement Result\tnm{a}\label{tab:measure}}
\footnotesize 
\centering
\begin{tabular}{ccccc||ccccc}
\hline\hline
SN Name & Type & $\lambda6150$\AA & O I $\lambda$7774\AA\  & Line Depth 
& SN Name & Type & $\lambda6150$\AA & O I $\lambda$7774\AA\  & Line Depth \\
 & & Depth & Depth & Ratio\tnm{b} & 
 & & Depth & Depth & Ratio\tnm{b} \\
\hline\hline
SN1989M & SN Ia & 0.703 & 0.264 & 2.67 &
SN1994D & SN Ia & 0.553 & 0.218 & 2.539 \\
SN1994S & SN Ia & 0.615 & 0.305 & 2.015 & 
SN1995D & SN Ia & 0.597 & 0.36 & 1.657 \\
SN1999gd & SN Ia & 0.687 & 0.271 & 2.535 &
SN1999gh & SN Ia & 0.711 & 0.417 & 1.706 \\
... & ... & ... & ... & ... & ... & ... & ... & ... & ... \\
SN2003gq & SN Ia-02cx & 0.165 & 0.275 & 0.6 &
SN2005hk & SN Ia-02cx & 0.187 & 0.246 & 0.759 \\
SN2008A & SN Ia-02cx & 0.199 & 0.27 & 0.738 &
SN1997br & SN Ia-91T & 0.144 & 0.153 & 0.94 \\
SN1991bg & SN Ia-91bg & 0.58 & 0.622 & 0.932 &
SN1999da & SN Ia-91bg & 0.648 & 0.673 & 0.963 \\
SN2001ex & SN Ia-91bg & 0.688 & 0.711 & 0.968 &
SN2002cf & SN Ia-91bg & 0.647 & 0.622 & 1.04 \\
... & ... & ... & ... & ... & ... & ... & ... & ... & ... \\
SN1998es & SN Ia-99aa & 0.432 & 0.279 & 1.546 &
SN1999aa & SN Ia-99aa & 0.432 & 0.259 & 1.668 \\
SN1999dq & SN Ia-99aa & 0.522 & 0.321 & 1.63 &
SN2001eh & SN Ia-99aa & 0.527 & 0.263 & 2.005 \\
... & ... & ... & ... & ... & ... & ... & ... & ... & ... \\
SN2000cx & SN Ia-pec & 0.384 & 0.199 & 1.932 &
SN1990I & SN Ib & 0.346 & 0.238 & 1.456 \\
SN1999dn & SN Ib & 0.311 & 0.15 & 2.074 &
SN2004dk & SN Ib & 0.268 & 0.148 & 1.814 \\
... & ... & ... & ... & ... & ... & ... & ... & ... & ... \\
PTF12hvv & SN Ib/c & 0.202 & 0.479 & 0.421 &
SN1999ex & SN Ib/c & 0.366 & 0.193 & 1.899 \\
SN2009jf & SN Ib/c & 0.224 & 0.4 & 0.559 &
SN2013ge & SN Ib/c & 0.316 & 0.438 & 0.721 \\
SN1991N & SN Ic & 0.038 & 0.167 & 0.23 &
SN1994I & SN Ic & 0.241 & 0.311 & 0.775 \\
SN2002ap & SN Ic & 0.134 & 0.454 & 0.295 &
SN2004aw & SN Ic & 0.173 & 0.359 & 0.481 \\
... & ... & ... & ... & ... & ... & ... & ... & ... & ... \\
PTF11img & SN Ic-BL & 0.193 & 0.103 & 1.872 &
SN2012ap & SN Ic-BL & 0.119 & 0.25 & 0.477 \\
... & ... & ... & ... & ... & ... & ... & ... & ... & ... \\
\hline\hline
\label{mresult}
\end{tabular}
\begin{tablenotes}
\item\tnm{a}{The table is abridged. The full table is available online.}
\item\tnm{b}{``Line Depth Ratio'' here means the ratio of  $\lambda6150$\AA\ depth to the O I $\lambda$7774\AA\ depth, i.e. $a{\rm (\lambda6150\AA)} / a{\rm (O\ I\ \lambda 7774\AA)}$}
\end{tablenotes}
\end{table*}

\subsubsection{Statistics of SNe Ia Line Depth Properties}

\begin{figure*}[!htp]
 \centering
  \includegraphics[width=0.49\linewidth]{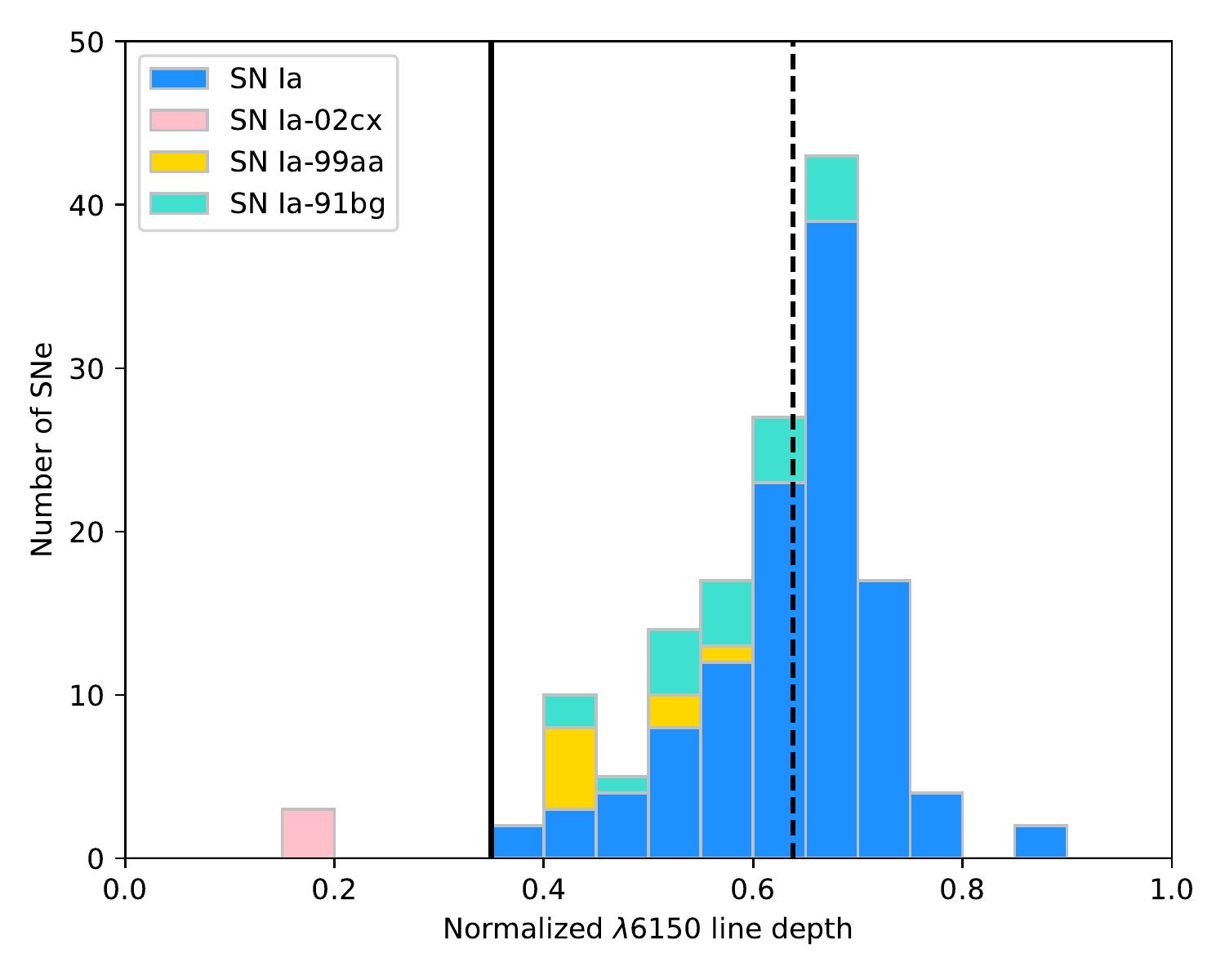}
  \includegraphics[width=0.49\linewidth]{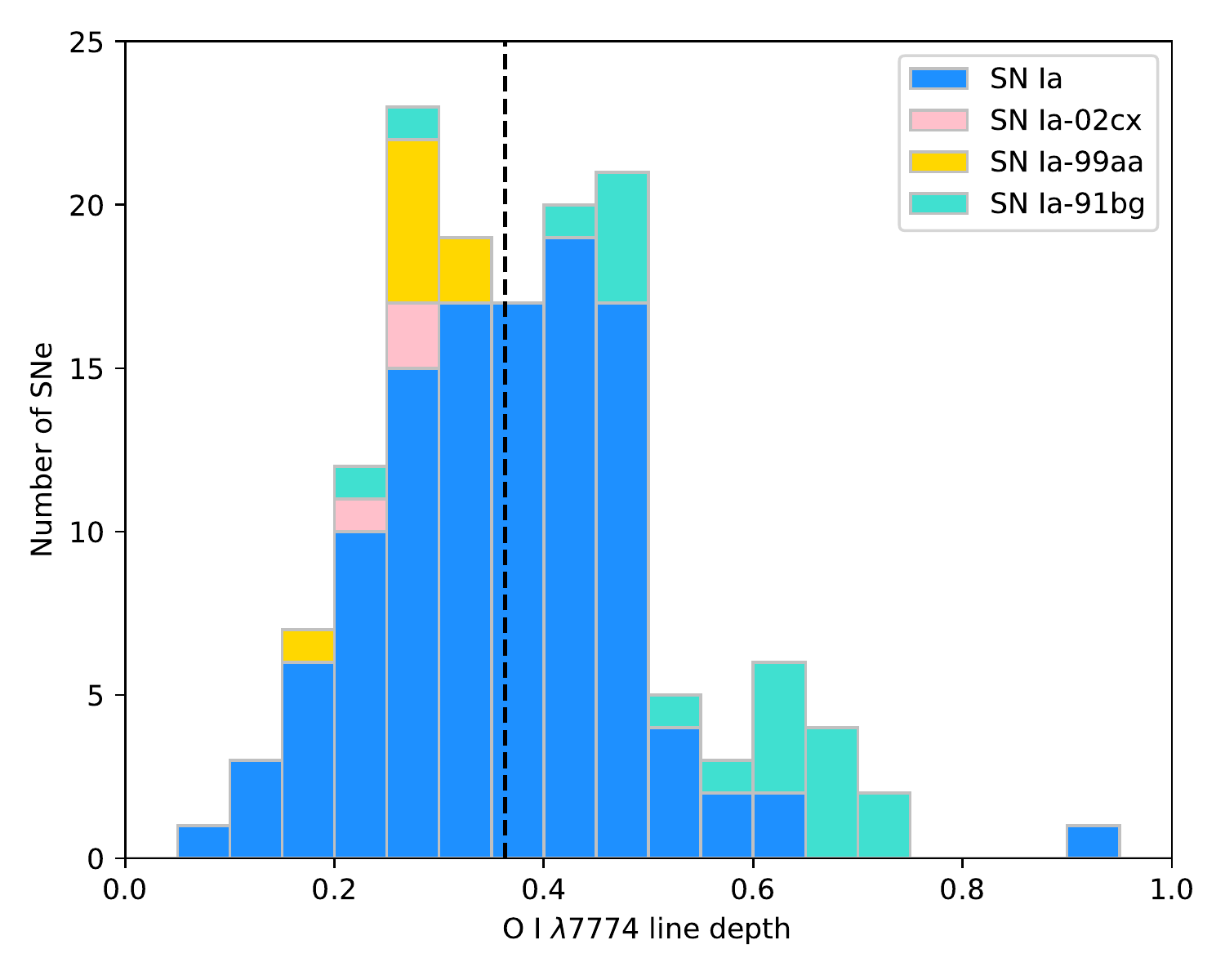}
  \caption{\footnotesize {\it Left}: Histogram of SNe Ia Si II $\lambda$6355\AA\ line depth distribution. The black vertical solid line represents $a{\rm (\lambda6150\AA)}=0.35$, while the dashed line represent the average for normal SNe Ia (blue bars) Si II $\lambda$6355\AA\ line depth (0.638).\quad {\it Right}: Histogram of the SN Ia O I $\lambda$7774\AA\ line depth distribution. The dashed line represents the average for normal SN Ia (blue bars) O I $\lambda$7774\AA\ line depth (0.363). Data in each bin are stacked together.} 
  \label{iahist}
\end{figure*}

In order to study the line depth distribution properties of SNe Ia in our sample, we plot two histograms of SN Ia Si II $\lambda$6355\AA\ and O I $\lambda$7774\AA\ line depth distributions respectively in the left and right panel of Figure \ref{iahist}. 
Since we have only a single example of the 1991T-like subclass (SN1997br) and one Ia-pec (SN2000cx), these are excluded from this analysis.

From the left plot in Figure \ref{iahist}, we can see that 141 of 144 (97.9\%) Type Ia SNe show a prominent Si II $\lambda$6355\AA\ absorption with $a{\rm (\lambda6150\AA)}>0.35$, except for the 3 SNe Ia-2002cx objects. The mean value for normal SNe Ia Si II $\lambda$6355\AA\ depth is 0.638, represented by the dashed vertical line on the plot, with a standard deviation of 0.089. Generally, peculiar SNe Ia have shallower Si II $\lambda$6355\AA\ lines than normal ones. SNe Ia-1991bg and Ia-1999aa are mostly located on the left of the dashed vertical line, while Ia-2002cx objects Si II $\lambda$6355\AA\ depths are even less than 0.2. The single 1991T-like object also lies below $a{\rm (\lambda6150\AA)}<0.35$, while SN2000cx is not an outlier ($a{\rm (\lambda6150\AA)}=0.384$).

In the right panel of Figure \ref{iahist}, we can find SNe Ia in almost every bin of the O I $\lambda$7774\AA\ line depth from 0.05 to 0.75. This phenomenon may indicate a diversity of O I optical depths in SN Ia photospheres. The dashed vertical line represents the mean value for normal SNe Ia O I $\lambda$7774\AA\ line depth (0.363). 
However, it is obvious that the Ia-1991bg and Ia-1999aa object distributions are separated in this figure. The majority of SNe Ia-1991bg are located on the right of the dashed line, except for two outliers (SN2006ke and SN2008bt), indicating a larger O I $\lambda$7774\AA\ line depth (mean: 0.551), while all the SNe Ia-1991aa are on the left (mean: 0.270). This could be related to the temperature sequence proposed by \citet{nugent95} extending from hot 1991T-like events to cool SNe Ia-1991bg. Three SNe Ia-2002cx also show shallow O I $\lambda$7774\AA\ line depth ($a{\rm (O\ I\ \lambda 7774\AA)}<0.3$). 


\subsubsection{Statistics of SN Ib and Ic Line Depth Properties}

Due to the low fraction of SNe Ib and Ic events among all Type I SNe, it is hard to obtain a large sample of SNe Ib\&c peak spectra to analyze. Therefore, the statistical work we carry out will be relatively limited. However, it is still feasible to find some distinction between SNe Ib and Ic based on their line depth features.

In the left panel of Figure \ref{ibchist} we present the histogram of SN Ib and Ic $\lambda$6150\AA \ line depth distribution. Except for a SN Ib outlier (PTF10feq) and a SN Ib/c outlier (SN1999ex), all of the objects in this plot are located on the left of the vertical solid black line at $a{\rm (\lambda6150\AA)}=0.35$, which is also the left boundary of normal SNe Ia we discussed above. SNe Ib $\lambda$6150\AA \ line depths are overall slightly larger than regular SNe Ic with significant overlap. The distribution range of SNe Ic-BL $\lambda$6150\AA \ line depth is consistent with that of regular SNe Ic.

The histogram in the middle panel of Figure \ref{ibchist} shows the SN Ib and Ic O I $\lambda$7774\AA\ line depth distributions. It is obvious in this subplot that SNe Ib show a generally shallower O I $\lambda$7774\AA\ absorption than regular SNe Ic. The phenomenon we revealed here is in accord with several previous studies \citep[for example,][and Fremling et al., in preparation]{2001AJ....121.1648M, 2016ApJ...827...90L}. SNe Ic-BL O I $\lambda$7774\AA\ line depths are somewhat lower than regular SNe Ic, i.e. closer to SNe Ib.

Since a small number of SNe Ib and regular SNe Ic are mixed in the central subplot, to distinguish them from each other, we adopt the line depth ratio of $\lambda$6150\AA \ to $\lambda$7774\AA, i.e. $a{\rm (\lambda6150\AA)} / a{\rm (O\ I\ \lambda 7774\AA)}$, as another index. The right panel of Figure \ref{ibchist} shows the histogram of SN Ib and Ic line depth ratio distribution. From this subplot we can find that SNe Ib and regular SNe Ic are separated almost perfectly on the two sides of the vertical solid line at $a{\rm (\lambda6150\AA)} / a{\rm (O\ I\ \lambda 7774\AA)}=1$, except for a SN Ic outlier, namely PTF12gzk. The line depth ratio of this object is only slightly above 1 (1.025), which is less than all the SNe Ib we analyze in this work. 
This result demonstrates that the line depth ratio could be used as a powerful index to distinguish the two types of SNe, though this index may not work well for SNe Ic-BL. Since Ic-BL can be easily distinguished from regular SNe Ib\&c by their extreme expansion velocities, leading to prominent line widths in their spectra, this is not a major shortcoming of the method.


\subsection{Quantitive Classification Criteria of Type I SNe}

Based on the line depth measurements and the statistical analysis reported above, here we intend to present quantitive classification criteria for Type I SNe. 



\begin{figure*}[!h]
 \centering
  \includegraphics[width=0.33\linewidth]{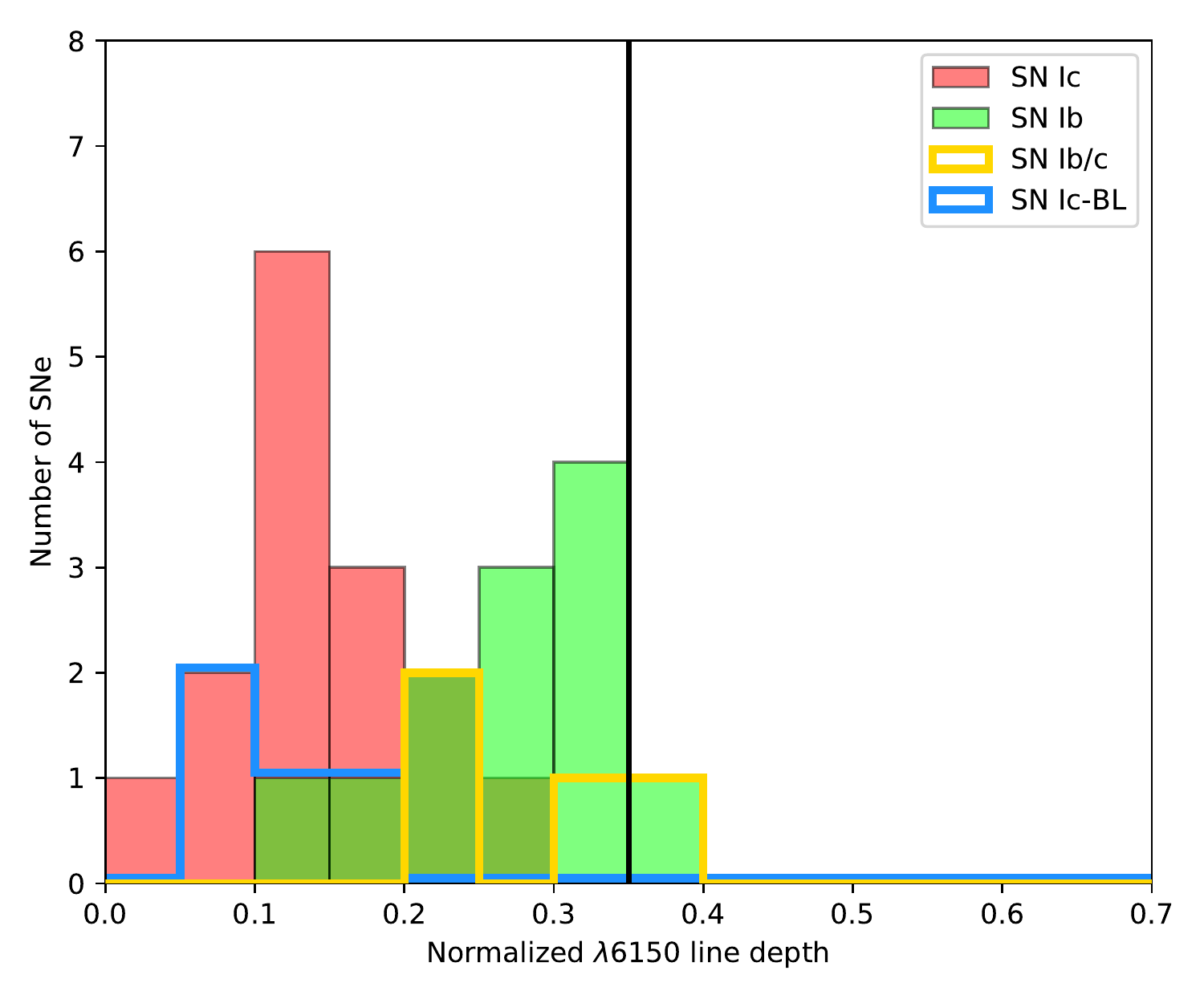}
  \includegraphics[width=0.33\linewidth]{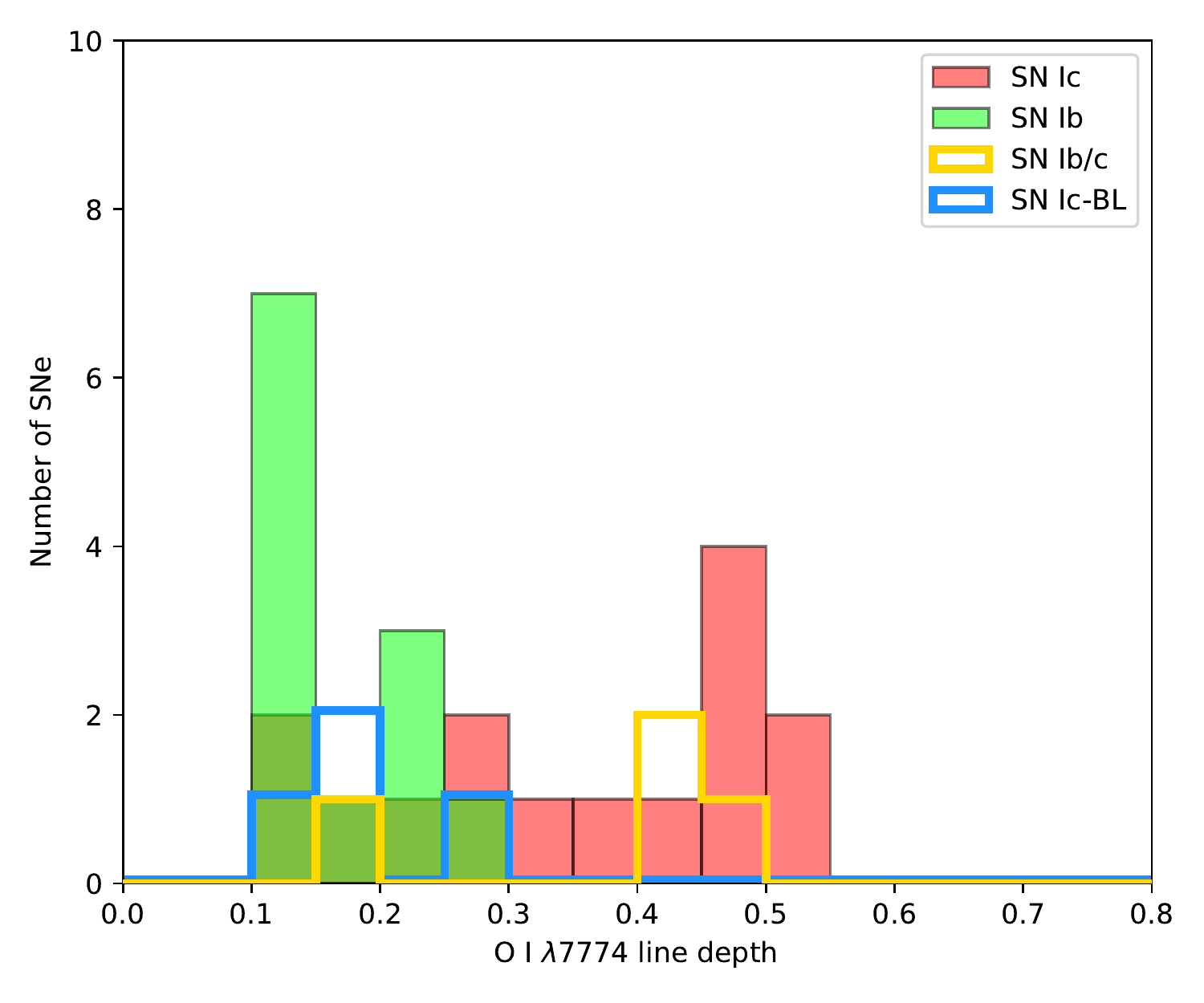}
  \includegraphics[width=0.33\linewidth]{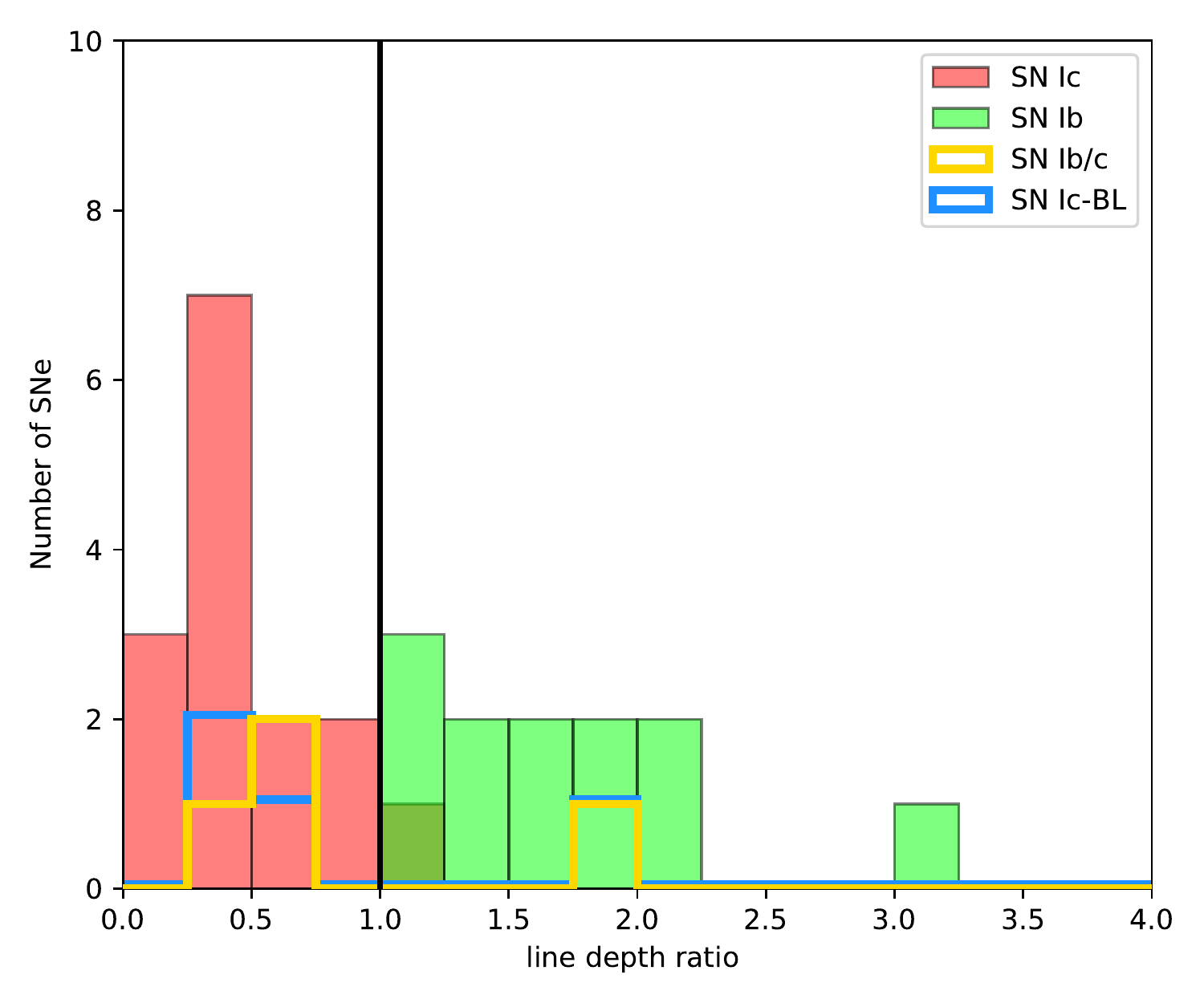}
  \caption{\footnotesize {\it Left}: Histogram of SNe Ib\&c's $\lambda$6150\AA \ line depth distribution. The vertical solid black line represents $a{\rm (\lambda6150\AA)}=0.35$, which is the left boundary of normal SNe Ia's $\lambda$6150\AA \ line depth.
  \quad {\it Middle}: Histogram of SNe Ib\&c's O I $\lambda$7774\AA\ line depth. 
  \quad {\it Right}: Histogram of SNe Ib\&c's line depth ratio of $\lambda$6150\AA \ to $\lambda$7774\AA\ . The vertical solid black line represents $a{\rm (\lambda6150\AA)} / a{\rm (O\ I\ \lambda 7774\AA)}=1$, which is the boundary shared by regular SNe Ib and regular SNe Ic. 
    } 
  \label{ibchist}
\end{figure*}

We have shown that all the regular SNe Ia, Ia-1991bg and Ia-1999aa events show prominent absorptions around $\lambda$6150\AA\ whose depths are larger than 0.35, while SNe Ib and Ic $\lambda$6150\AA\ depths are below this threshold. Therefore, it encourages us to adopt $a{\rm (\lambda6150\AA)}$ as the first index to distinguish SNe Ia from regular SNe Ib and Ic. To show the distinction in OI $\lambda$7774\AA\ line depth between SNe Ib and SNe Ic clearly, we decide to use the line depth ratio of $\lambda$6150\AA \ to OI $\lambda$7774\AA\ as the second index (Figure \ref{ibchist}). We introduce the following set of quantitative classification criteria for the majority of Type I SNe as formulae: 

1. SNe Ia (including normal Ia, Ia-1991bg and Ia-1999aa): 

\begin{equation}
a{\rm (\lambda6150\AA)}>0.35
\end{equation}

2. SNe Ib: 
\begin{equation}
\begin{aligned}
a{\rm (\lambda6150\AA)} &<0.35; \\ 
a{\rm (\lambda6150\AA)} &/ a{\rm (O\ I\ \lambda 7774\AA)} >1
\end{aligned}
\end{equation}

3. SNe Ic (except for Ic-BL):
\begin{equation}
\begin{aligned}
a{\rm (\lambda6150\AA)} &<0.35; \\ 
a{\rm (\lambda6150\AA)} &/ a{\rm (O\ I\ \lambda 7774\AA)} <1
\end{aligned}
\end{equation}

\begin{figure*}[ht]
  \centering
  \includegraphics[width=0.9\linewidth]{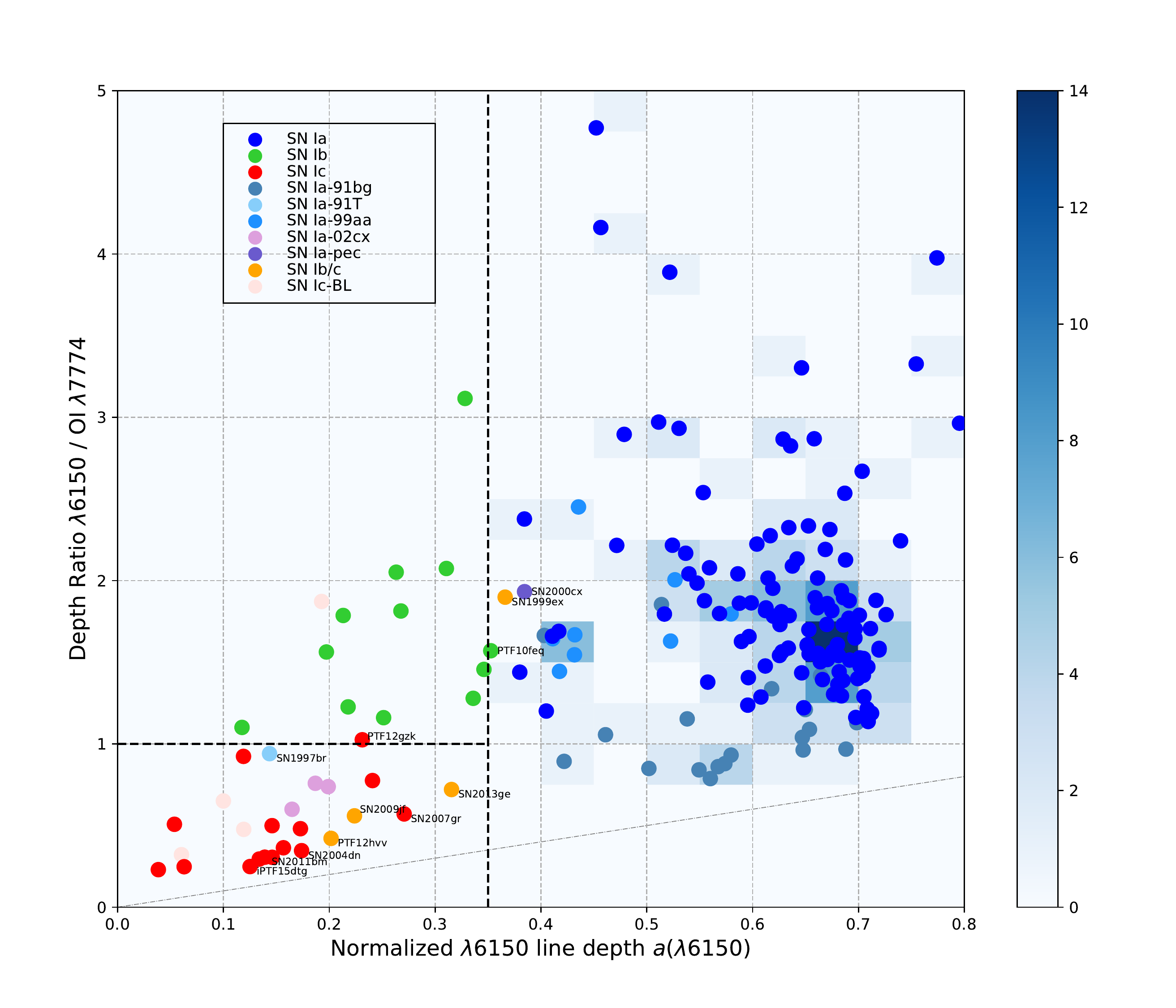}
  \caption{\footnotesize $\lambda$6150\AA\ line depth versus line depth ratio of $\lambda$6150\AA \ to OI $\lambda$7774\AA\ of 181 Type I SNe. From this plot we can see the majority of SNe Ia (various kinds of blue dots; representing normal SNe Ia, Ia-1991bg, Ia-1999aa respectively) are located on the right of the vertical dashed line at $a{\rm (\lambda6150\AA)}=0.35$, while regular SNe Ib (green dots) are on the upper-left part of the plot. Regular SNe Ic (red dots) as well as Ia-2002cx objects (violet dots) show a relatively shallower $\lambda$6150\AA \ than normal SNe Ia but deeper OI $\lambda$7774\AA\ absorption, so they are co-located with SNe Ic in the lower-left corner. All the names of SNe Ib/c (golden dots), the Ib outlier (PTF10feq), Ia-1991T (SN1997br as the light skyblue dot), Ia-pec (SN2000cx as the purple dot) as well as several SNe Ic are annotated on the plot. The background of the plot is a density map for normal SNe Ia, Ia-1991bg and Ia-1999aa, with the relevant color bar on the right.}
  \label{majorfig}
\end{figure*}

To show these criteria schematically, we plot a scatter diagram (Figure \ref{majorfig}) to show all the 181 Type I SNe distribution of $\lambda$6150\AA\ line depth and line depth ratio of $\lambda$6150\AA \ to OI $\lambda$7774\AA. From the figure we can see that all the normal SNe Ia, SNe Ia-1991bg and SNe Ia-1999aa are located on the right of vertical black dashed line at $a{\rm (\lambda6150\AA)}=0.35$. SNe Ib mainly lie in the upper-left part of the diagram, except for a narrow outlier PTF10feq ($a{\rm (\lambda6150\AA)}=0.353$). Regular SNe Ic are situated in the bottom left corner with $a{\rm (\lambda6150\AA)}<0.35$ but $a{\rm (\lambda6150\AA)} / a{\rm (O\ I\ \lambda 7774\AA)}<1$ except for PTF12gzk. Since PTF12gzk is reported to show high ejecta expansion velocities in peak spectra \citep[e.g.][]{2012ApJ...760L..33B,2016ApJ...832..108M} resembling SN Ic-BL, its spectral line depth properties may have slight differences from regular SNe Ic.
 SNe Ia-2002cx objects locate in the same region as regular SNe Ic, while SNe Ic-BL seem to show up irregularly on the left of the vertical black dashed line. A single Ia-91T event is also an outlier.

\subsection{Application and Discussion}
\label{sec:app}
The era of rapid development of observational techniques and instruments and international cooperation results in many SN discoveries, as well as many novel and strange SN subclasses. Among Type I SNe, the intermediate subtype between SN Ib and Ic, namely SN Ib/c, attracts our attention. 

It is commonly reckoned that both SNe Ib and Ic originate from core collapse of massive stars that were stripped of their outermost hydrogen envelopes. 
 For most conditions, SNe Ib are distinctive from SNe Ic due to helium features in their spectra around maximum light. However, the discovery of Type I SNe with transitional spectroscopic features between Type Ib and Ic SNe \citep[e.g. SN1999ex, ][]{2002AJ....124..417H, 2002AJ....124.2100S} indicates the difficulties of distinguishing these two types clearly and accurately.

Since the basic purpose of SN classification is to differentiate these events to several groups by their shared characters and distinctive features, we prefer to give a clear identifications to intermediate or unclear SNe Ib/c rather than leave them under an umbrella group. As we have already introduced a set of quantitative classification criteria for Type I SNe, it would be a feasible application and examination of our criteria to give clear identifications to these so-called SNe Ib/c. 

Among our 4 SNe Ib/c objects, 3 (PTF12hvv, SN2009jf, SN2013ge) are located in the region of regular SNe Ic. PTF12hvv does not show conspicuous helium absorption in its optical spectrum, therefore it could be reclassified as a Type Ic SN. Similarly, SN2013ge \citep{Drout2016} and SN 2009jf \citep{2011MNRAS.416.3138V} also show weak helium features, especially at both $\lambda$6678\AA\ and $\lambda$7065\AA\ , resembling SNe Ic. 


Another SN Ib/c in our sample, SN1999ex, is located beyond but quite close to the right boundary of the SNe Ib region, with $a{\rm (\lambda6150\AA)}= 0.366$, which is quite similar to PTF10feq. With the notable He I $\lambda\lambda\lambda$5876\AA\ 6678\AA\ 7065\AA\ features in its peak spectra, we have reason to believe this object can be reclassified as a Type Ib SN.

From all above we can carefully draw a conclusion that the majority of Type I SNe could be quantitatively classified according to their peak spectra. By measuring the line depth of $\lambda$6150\AA \ and OI $\lambda$7774\AA\ absorption around maximum light, we can simply give their location on the diagram and get their classification, with the only exception being the peculiar Ia-2002cx, Ic-BL and possibly Ia-91T subtypes. Meanwhile, there is no more need to identify Helium feature in their spectra to give an exact classification of Type I CCSNe, which may help to enhance the accuracy of determining a Type Ib or Ic classification. 

Though a more detailed classification of a Type I SN may benefit from comparison and matching between the candidate spectrum and some well-calibrated template, or even better, a spectroscopic model \citep[e.g.][]{Hachinger2012}, this simple and interactive line-depth classification method may give a quantitative way to classify these SNe, and thus could play an important role in the rapidly developing era of time-domain astronomy. Meanwhile, the physical implication of $a{\rm (\lambda6150\AA)} / a{\rm (O\ I\ \lambda 7774\AA)}$ we introduced is not clear. This would be a good and natural question to target with physical models. 

\section{Summary}
We have carried out a set of interactive $\lambda$6150\AA\ and O I $\lambda$7774\AA\ absorption line depth measurements on 181 Type I SN peak spectra. The Type I SN sample in this paper includes 146 SNe Ia (consisting of 133 normal Ia, 1 Ia-1991T, 21 Ia-1991bg, 8 Ia-1999aa and 3 Ia-2002cx), 12 SNe Ib, 19 SNe Ic (including 5 SNe Ic-BL) and 4 intermediate SN Ib/c. All the SN Ia objects we analyze here are from the low-redshift ($z<0.1$) BSNIP sample, while the SNe Ib and Ic we study here are acquired from WISeREP, originally from various publications, including 6 unpublished PTF and iPTF CCSNe. All the spectra are taken within 5 days of the SN maximum dates. For SNe Ia the dates of maximum are defined in B-band while V-band or g-band is typically used for SNe Ib and Ic, due to previous studies biases. We found that the majority of Type Ia SNe (excluding Ia-2002cx) show prominent $\lambda$6150\AA\ absorption with depths greater than 0.35, while SNe Ib and Ic do not exceed this threshold. SNe Ib generally show a shallower O I $\lambda$7774\AA\ absorption than regular SNe Ic, resulting in a line depth ratio of $\lambda$6150\AA\ to O I $\lambda$7774\AA\ greater than 1. This value is below 1 for the regular SNe Ic in our dataset. Therefore we generalize a set of quantitative classification criteria for Type I SNe. These criteria are based on line depth measurements of $\lambda$6150\AA\ and O I $\lambda$7774\AA\ absorption regions in SN peak spectra and can distinguish all major classes of Type I SNe, except for peculiar SNe Ia-2002cx and Ic-BL events. 
We apply this differentiating method to 4 transitional or unclear SNe Ib/c spectra. We find the classification results we derive are in high consistency with some previous classifications by identifying helium features in SN spectra. 
The physical meaning of the line depth ratio of $\lambda$6150\AA\ to O I $\lambda$7774\AA\ remains in need of an explanation.

\acknowledgments

We thank R. Ellis, M. Sullivan, P. E. Nugent and A. V. Filippenko for use of PTF follow-up data in advance of publication. We thank S. Prentice for his helpful dataset of CCSN maximum dates in multicolor. We thank P. Nugent and J. Sollerman for useful discussions and comments. 
This research made use of Astropy, a community-developed core Python package for Astronomy \citep{2013A&A...558A..33A}; matplotlib, a Python library for publication quality graphics \citep{Hunter:2007} and NumPy \citep{van2011numpy}. We thank the NASA's Astrophysics Data System (ADS) for abstract and bibliography services; the SNDB for Type Ia SNe data; The Open Supernova Catalog for SN spectra query. We specially thank WISeREP for access to SN data.

Part of this work was done during FS's visit to the Weizmann Institute of Science (WIS) and we thank WIS for supporting this visit. FS is also supported by the Undergraduate Research Training Program of Peking University.
AGY is supported by the EU via ERC grants No. 307260 and 725161, the Quantum Universe I-Core program by the Israeli Committee for Planning and Budgeting, and the ISF; a Binational Science Foundation "Transformative Science" grant and by a Kimmel award.


\bibliography{QCC_SNeI.bib}




\end{document}